\documentclass[twocolumn]{aastex631}
\usepackage{amsmath}

\newcommand{\angstrom}{\text{\normalfont\AA}}

\shorttitle{Stripe 82X Catalog Release}
\shortauthors{LaMassa et al.}

\graphicspath{{./}{figures/}}

\begin{document}

\title{Stripe 82X Data Release 3: Multiwavelength Catalog with New Spectroscopic Redshifts and Black Hole Masses}

\correspondingauthor{Stephanie LaMassa}
\email{slamassa@stsci.edu}

\author[0000-0002-5907-3330]{Stephanie LaMassa}
\affiliation{Space Telescope Science Institute
3700 San Martin Dr.,
Baltimore, MD, 21218, USA}

\author[0000-0003-2196-3298]{Alessandro Peca}
\affiliation{Department of Physics,
  University of Miami, 
  Coral Gables, FL 33124, USA}

\author[0000-0002-0745-9792]{C. Megan Urry}
\affiliation{Department of Physics, Yale University,
  New Haven, CT 06520, USA}
\affiliation{Yale Center for Astronomy and Astrophysics,
  New Haven, CT 06520, USA}

\author[0000-0003-0489-3750]{Eilat Glikman}
\affiliation{Department of Physics, Middlebury College, Middlebury, VT 05753, USA}

\author[0000-0001-8211-3807]{Tonima Tasnim Ananna}
\affiliation{Department of Physics and Astronomy, Wayne State University, Detroit, MI 48202, USA}

\author[0000-0002-5504-8752]{Connor Auge}
\affiliation{Institute for Astronomy, University of Hawaii, Honolulu, HI, 96822, USA}

\author[0000-0002-2115-1137]{Francesca Civano}
\affiliation{ Astrophysics Science Division, NASA Goddard Space Flight Center, Greenbelt, MD 20771, USA}

\author[0000-0002-2525-9647]{Aritra Ghosh}
\affiliation{Department of Astronomy, Yale University, New Haven, CT 06520, USA}
\affiliation{Yale Center for Astronomy and Astrophysics, New Haven, CT 06520, USA}

\author[0000-0002-5537-8110]{Allison Kirkpatrick}
\affiliation{Department of Physics and Astronomy, University of Kansas, Lawrence, KS 66045, USA}

\author[0000-0002-7998-9581]{Michael J. Koss}
\affiliation{Eureka Scientific,
  2452 Delmer Street Suite 100, 
  Oakland, CA 94602-3017, USA}

\author[0000-0003-2284-8603]{Meredith Powell}
\affiliation{Leibniz-Institut f\"ur Astrophysik Potsdam, Postdam, Germany}
\affiliation{Schwarzschild Fellow}

\author[0000-0001-7116-9303]{Mara Salvato}
\affiliation{Max-Planck-Institut f\"ur extraterrestrische Physik, Garching, Germany}
\affiliation{Exzellenzcluster ORIGINS, Boltzmannstr.  Garching, Germany}

\author[0000-0002-3683-7297]{Benny Trakhtenbrot}
\affiliation{School of Physics and Astronomy,
  Tel Aviv University, 
  Tel Aviv 69978, Israel}

\begin{abstract}
  We present the third catalog release of the wide-area (31.3 deg$^2$) Stripe 82 X-ray survey. This catalog combines previously published X-ray source properties with multiwavelength counterparts and photometric redshifts, presents 343 new spectroscopic redshifts, and provides black hole masses for 1297 Type 1 Active Galactic Nuclei (AGN). With spectroscopic redshifts for 3457 out of 6181 Stripe 82X sources, the survey has a spectroscopic completeness of 56\%. This completeness rises to 90\% when considering the contiguous portions of the Stripe 82X survey with homogeneous X-ray coverage at an optical magnitude limit of $r<22$. Within that portion of the survey, 23\% of AGN can be considered obscured by being either a Type 2 AGN, reddened ($R-K > 4$, Vega), or X-ray obscured with a column density $N_{\rm H} > 10^{22}$ cm$^{-2}$. Unlike other surveys, there is only a 18\% overlap between Type 2 and X-ray obscured AGN. We calculated black hole masses for Type 1 AGN that have SDSS spectra using virial mass estimators calibrated on the H$\beta$, \ion{Mg}{2}, H$\alpha$, and \ion{C}{4} emission lines. We find wide scatter in these black hole mass estimates, indicating that statiscal analyses should use black hole masses calculated from the same formula to minimize bias. We find that the AGN with the highest X-ray luminosities are accreting at the highest Eddington ratios, consistent with the picture that most black hole mass accretion happens in the phase when the AGN is luminous ($L_{\rm 2-10 keV} > 10^{45}$ erg s$^{-1}$).

\end{abstract}


\section{Introduction}

Multi-wavelength surveys give us a view into the cosmic evolution of supermassive black holes and the galaxes in which they live. When a steady supply of material accretes onto a supermassive black hole, the centers of these galaxies shine brightly as Active Galactic Nuclei (AGN). Studying AGN from the early Universe to the present day via surveys allows us to measure the growth history of supermassive black holes \citep{soltan1982}.

AGN emit across the electromagnetic spectrum and can thus be identified in surveys at any wavelength \citep[see][for a review]{hickox2018}. Over three-quarters of a million AGN have been discovered in optical surveys like the Sloan Digital Sky Survey \citep[SDSS;][]{hao2005,lyke2020}, though optical selection favors unobscured Type 1 AGN where we have a direct view of the bright ultraviolet and optical emission from the accretion disk and broad line region. Most AGN are obscured \citep{ueda2014,aird2015,buchner2015,ananna2019} where our line-of-sight to the AGN central engine is blocked by dust and gas \citep{antonucci1993,urry1995,netzer2015}, causing the ultraviolet and optical light to be attenuated or extinguished, leaving only narrow lines visible in the spectrum (Type 2 AGN). Infrared selection is a powerful probe to identify obscured AGN \citep{donley2012,mateos2012}, but it is challenging to differentiate infrared emission caused by accretion disk heated dust versus vigorous star formation \citep[e.g.,][]{mendez2013,ichikawa2017}. X-rays, produced by inverse Compton scattering of accretion disk photons to X-ray energies, provide the cleanest probe to identify AGN as star-forming processes do not produce X-ray emission as luminous as AGN \citep{brandt2005}. However, the most heavily obscured AGN (i.e., the Compton-thick population) suppresses the observed X-ray emission, making this population challenging to identify in X-ray surveys that are flux limited \citep[see, e.g., Figure 2 of][]{koss2016}. Additionally, X-ray surveys are biased against low luminosity AGN, including those in low mass, low metallicity galaxies \citep[e.g.,][]{cann2020}. Radio surveys are powerful for identifying AGN \citep{white2000, hodge2011,helfand2015} and are relatively insensitive to obscuration \citep{webster,glikman2004}, though star forming galaxies dominate the radio galaxy population at flux densities below $S_{\rm 1.4 GHz}\sim$ 100 $\mu$Jy \citep[e.g.,][]{bonzini2013}. Still, current and planned sensitive radio surveys like MeerKat \citep{jonas2016} and, its successor, the Square Kilometer Array \citep{dewdney2009} will increase the census of AGN and shed light on their radio properties. Since any one selection method has both strengths and weaknesses, combining data from multi-wavelength surveys is essential to reveal the full AGN population across the Universe.

In addition, different survey strategies preferentially identify different AGN populations. Deep, pencil-beam surveys that cover areas under one square degree, like the Great Observatories Origins Deep Survey \citep[GOODS,][]{giavalisco2004} within the {\it Chandra} Deep Field South field \citep{giacconi2002,lehmer2005,luo2008,luo2017,xue2011}, unveil the faintest objects but do not adequately sample objects that have a low space density. They are also subject to cosmic variance. Such deep surveys have returned an impressive array of the most distant galaxies identified to date with the $\sim0.0125$ deg$^2$ JWST Deep Extragalactic Survey \citep[$z \sim 10-13$;][]{rieke2023,eisenstein2023,curtis-lake2023,bunker2023} and $\sim0.01$ deg$^2$ JWST Cosmic Evolution Early Release Science (CEERS) survey \citep[$z \sim 7 - 10$;][]{finkelstein2023,arrabalharo2023,fujimoto2023}. However, the deepest X-ray survey (the 7 Ms {\it Chandra} Deep Field South) detects AGN only as distant as $z \sim 5$ \citep{luo2017}, though more distant AGN ($z > 6$), that were first discovered in optical surveys, have been detected in wide-field X-ray surveys like the all-sky eROSITA survey  \citep{predehl2021,medvedev2020,medvedev2021,wolf2021,wolf2023}. The most distant X-ray AGN identified thus far at $z = 10.3$ was a serendipitous discovery from the Abell 2744 cluster field observed with JWST \citep{goulding2023}, leveraging gravitational lensing of the cluster to boost the X-ray signal from the background AGN to make it detectable \citep{bogdan2023}. Such small number statistics in the high-redshift AGN population lead to uncertainties in the role of AGN to cosmic reionization, where estimates of the AGN contribution range from significant \citep[e.g.,][]{madau2015} to neglible \citep[e.g.,][]{jiang2022}, with perhaps a dichotomy where AGN are significant players in helium reionization but not hydrogen \citep{yung2021}.

Moderate-area, moderate-depth surveys of several square degrees sacrifice depth to cover a greater area and better sample the typical AGN population compared with deeper surveys. For instance, from surveys like {\it Chandra} COSMOS Legacy covering 2.2 deg$^2$ \citep{civano2016,marchesi2016a}, we can measure how the space density of moderate luminosity (10$^{43}$ erg s$^{-1} < L_{\rm 2-10 keV} < 10^{44}$ erg s$^{-1}$) AGN evolve from $z\sim 4$ to the current Universe \citep{marchesi2016b}. However, rare objects, like high luminosity AGN (i.e., quasars, $L_{\rm 2-10 keV} > 10^{45}$ erg s$^{-1}$) are still under represented in moderate-area, moderate-depth surveys. Though rare, these objects are important as they represent a relatively short-lived phase where the bulk of mass accretion onto supermassive black holes occur \citep{hopkins2009,treister2012}. The obscured quasar phase may also represent a critical time in the co-evolution of galaxies and black holes when the central engine is temporarily encased in dust and gas just prior to the expulsion of this material from AGN accretion disk winds \citep{sanders1988,hopkins2006}. 

Stripe 82X is a wide-area X-ray survey designed to discover obscured quasars that had previously been missing from the global census of black hole growth. Chosen to overlap the Stripe 82 legacy field in SDSS\citep{frieman2008}, Stripe 82X combines rich multi-wavelength coverage from the ultraviolet \citep[GALEX;][]{morrissey2007}, optical \citep[coadded SDSS catalogs;][]{annis2014,jiang2014,fliri2016}, near-infrared \citep[UKIDSS and VSH;][]{hewett2006,casali2007,lawrence2007}, mid-infrared \citep[{\it WISE} and {\it Spitzer};][]{wright2010,allwise,timlin2016,papovich2016}, far-infrared \citep[{\it Herschel};][]{viero2014}, and radio \citep[FIRST;][]{becker1995,hodge2011,helfand2015} with X-ray data from archival {\it Chandra} \citep{lamassa2013a} and {\it XMM-Newton} observations \citep{lamassa2013b} and dedicated {\it XMM-Newton} observing campaigns in AO-10 and AO-13 \citep[PI: C. M. Urry;][]{lamassa2013b,lamassa2016a}.

  The archival X-ray data in Stripe 82 are by nature heterogeneous in areal coverage and depth, while more homogeneous coverage is obtained with the dedicated {\it XMM}-AO10 and {\it XMM}-AO13 observing campaigns. The {\it XMM}-AO10 data consists of two $\sim$2.3 deg$^2$ patches of sky \citep{lamassa2013b}, reaching an X-ray full band flux limit of $F_{\rm 0.5-10 keV} \sim 9 \times 10^{-15}$ erg s$^{-1}$  cm$^{-2}$, with a half-area survey flux limit of $1.7 \times 10^{-14}$ erg s$^{-1}$ cm$^{-2}$.  The {\it XMM}-AO13 data represents the largest contiguous X-ray area in Stripe 82, with $\sim$15.6 deg$^2$ chosen to overlap pre-existing {\it Spitzer} and {\it Herschel} coverage. Stripe 82 {\it XMM}-AO13 has a flux limit of $F_{\rm 0.5 - 10 keV} = 6.7 \times 10^{-15}$ erg s$^{-1}$ cm$^{-2}$ and a half-area survey flux limit of $1.5 \times 10^{-14}$ erg s$^{-1}$ cm$^{-2}$. Combining the archival X-ray, {\it XMM} AO10, and {\it XMM} AO13 data, Stripe 82X covers 31.3 deg$^2$, with 6181 unique X-ray sources detected. It had an initial spectroscopic completeness of $\sim$30\% \citep[i.e., 1842 X-ray sources had spectroscopic redshifts;][]{lamassa2016a} from SDSS and other spectroscopic surveys (see the Appendix for a full list). Of these 6181 X-ray sources, secure multi-wavelength counterparts were identified for 6053, with photometric redshifts calculated for 5971 \citep{ananna2017}. A special SDSS-IV \citep{gunn2006,blanton2017} eBOSS \citep{smee2013,dawson2016} program targeted the Stripe 82X AO-13 field, providing spectroscopic redshifts for an additional 616 sources \citep{lamassa2019}.

In this third data release of the Stripe 82X catalog, we combine the catalogs from previous releases to produce a comprehensive user-friendly catalog, including updates to SDSS spectroscopic classifications when warranted (Section \ref{vet_sdss}), and report an additional 343 spectroscopic redshifts obtained from our dedicated ground-based observing campaigns with Palomar and Keck that have not been previously published (Section \ref{new_specz}). We identify an X-ray flux - optical magnitude parameter space where Stripe 82X is $>90\%$ spectoscropically complete ($F_{\rm 0.5-10 keV} \gtrsim 6 \times 10^{-15}$ erg s$^{-1}$ cm$^{-2 }$; $r < $22 AB) and comment on the obscured AGN demographics in this sector (Section \ref{obsc_agn_sec}). We fit the SDSS spectra of Stripe 82X Type 1 AGN to measure black hole masses where possible (Section \ref{bhmass_meas}) to include in this catalog and estimate the Eddington ratios of these objects. A forthcoming Stripe 82X catalog will be published that will include new archival {\it Chandra} and {\it XMM}-Newton observations within the Stripe 82 field, for a total Stripe 82 X-ray coverage of$\sim$55 deg$^2$ with $\sim$23000 objects detected (Stripe 82-XL; Peca et al., accepted.).

\section{Creating the Stripe 82X Data Release~3 Catalog}\label{cat_updates}
The current release of the Stripe 82X catalog combines the X-ray information in the \cite{lamassa2016a} catalog (Data Release 1) with the multi-wavelength counterpart identification and photometric redshifts\footnote{The photometric redshifts were calculated by fitting templates to the multi-wavelength spectral energy distributions (SEDs) of Stripe 82X sources.} released in the \citet{ananna2017} catalog (Data Release 2), as well as pre-existing spectrocopic redshifts from independent surveys, spectroscopic redshifts obtained through the special SDSS-IV eBOSS program \citep{lamassa2019}, and spectroscopic redshifts determined from dedicated ground-based observing campaigns with Palomar and Keck. This catalog also contains black hole masses calculated from fitting the SDSS spectra of Type 1 AGN in Stripe 82X which we discuss below in greater detail.

While a full description of the columns in this catalog is presented in the Appendix, we summarize here two additional columns not included in previous catalog releases. These columns facilitate identifying which sources are X-ray AGN, defined by having a 2-10 keV X-ray luminosity greater than 10$^{42}$ erg s$^{-1}$ \citep{brandt2005}. First, we use the observed X-ray count rate to estimate the $k$-corrected (rest-frame) X-ray hard band (2-10 keV) luminosity ($L_{\rm X}$) for all extragalactic objects with a spectroscopic redshift. The $k$-corrected luminosity is calculated as $L_{\rm k-corr} = L_{\rm observed} \times (1 + z)^{\Gamma -2}$, where $\Gamma$ is the spectral slope of the power law model used to convert X-ray count rates to fluxes. As discussed in \citet{lamassa2013a,lamassa2016a}, we assumed a model where $\Gamma$ = 2 for the soft band (0.5-2 keV) and $\Gamma$=1.7 for the hard and full (0.5-10 keV) bands to convert the X-ray count rate to observed X-ray flux.

In order for an X-ray source to be included in the Stripe 82X catalog, it had to be detected at the 4.5$\sigma$ level \citep[{\it Chandra};][]{lamassa2013a} or the 5$\sigma$ level \citep[{\it XMM-Newton};][]{lamassa2013b,lamassa2016a} in at least one of the X-ray energy bands and to be included in the Log$N$-Log$S$ distribution for that band.  Thus the flux measurement reported in the hard energy band is not necessarily significant and there may be no hard X-ray dection for some sources. We follow the procedure discussed in \citet{lamassa2019} to estimate $L_{\rm X}$. If an X-ray source is detected at a signficance level $\geq 4\sigma$ in the hard band, $L_{\rm X}$ is the $k$-corrected hard X-ray luminosity. If an X-ray source is not detected at that significance in the hard band, then the rest-frame 2-10 keV luminosity is estimated based on scaling the full band or soft band X-ray luminosity according to the X-ray spectral model cited above. If a source is detected at $\geq 4\sigma$ in the full band, the $k$-corrected full band luminosity is scaled by a factor of 0.665 to estimate $L_{\rm X}$. If an X-ray source is not detected at $\geq 4\sigma$ in the hard nor full band, then $L_{\rm X}$ represents the soft band flux multiplied by 1.27. $L_{\rm X}$ is reported as ``X-ray\_Lum'' in the catalog. If this value exceeds $10^{42}$ erg s$^{-1}$, the ``X-ray\_AGN'' column in the catalog is set to ``True.'' For intrinsic (absorption-corrected) X-ray luminosities based on fitting the X-ray spectra of the Stripe 82X AGN, we refer the reader to \citet{peca2023}.

In the analysis presented here, we use the X-ray luminosities estimated via their observed X-ray counts and assuming a spectral model as described above. While the X-ray luminosities derived via spectral modeling are more accurate, this information is only available for a subset of the full Stripe 82X sample. However, when calculating Eddington ratios in Section \ref{ledd_sec}, we limit our analysis to the subset of AGN with intrinsic X-ray luminosities calculated via spectral fitting and presented in \citet{peca2023}.

\subsection{Vetting SDSS Spectroscopic Classification}\label{vet_sdss}
Visual inspection of the SDSS spectra of Stripe 82X sources reveals that sometimes objects spectroscopically identified as ``GALAXY'' by the SDSS pipeline have broad lines that are not fitted by the spectral templates. Such objects are better classified as broadline AGN, or, following the SDSS naming convention, ``QSO.'' To systematically check for such mis-identified objects, we fitted the spectra of all 573 X-ray sources that the SDSS spectroscopic pipeline classified as a ``GALAXY.'' We used the Galaxy/AGN Emission-Line Analysis TOol \citep[GELATO;][]{gelato,hviding2022} to perform the spectral fitting and test whether significant broad lines are detected for the H$\alpha$, H$\beta$, \ion{Mg}{2}, or \ion{C}{3} lines. We performed this fitting regardless of whether the source is an X-ray AGN (i.e., $L_{\rm X} > 10^{42}$ erg s$^{-1}$) or not.

GELATO attempts to fit broad Gaussian components to these emission lines with a lower bound on the dispersion constrained to 500 km s$^{-1}$ \citep[corresponding to FWHM = 1200 km s$^{-1}$, which is the empirical dividing line between Type 1 and Type 2 AGN;][]{hao2005} and a higher bound constrained to 6500 km s$^{-1}$. If a broad Gaussian component is found, GELATO returns the fitted parameters for that component.

We perform some additional checks to vet these results to distinguish between true broad lines and spurious detections where noise in the spectrum could be fitted by a broadened Gaussian line. First, we require that the fitted dispersion values not be at the limits of 500 km s$^{-1}$ or 6500 km s$^{-1}$ because such values indicate that noise is being fitted. We confirmed via visual inspection that these values at the allowed limits of the fitting were cases where GELATO indeed was fitting noise in the spectrum. We also require that the signal-to-noise ratio (S/N) in the line flux and line equivalent width exceed 3. We then calculate the Amplitude-over-Noise (AoN) ratio for each line, which is the Amplitude of the Gaussian divided by the 3-sigma-clipped root-mean-square (RMS) of the model fit to the spectrum in the continuum on either side of the emisison line. We define the local continuum to start 5$\sigma$ away from the fitted line center (where $\sigma$ is the fitted Gaussian dispersion) with a 100 $\angstrom$ window (50 $\angstrom$ for H$\beta$ due to the nearby [\ion{O}{3}] doublet) on either side of the emission line. We consider a broad line to be detected if AoN $\geq$ 3. Finally, we visually inspect the results to identify any broad line detections that look spurious and any sources where we found that GELATO did not fit a broad component that is visible in the spectrum, leaving residuals between the fitted line and spectrum that are identifiable by eye. In total, we find that 39 objects out of 573 originally identified as a ``GALAXY'' from the SDSS pipeline have a significant broad line detection, so we reclassify these objects as a ``QSO'' in the Stripe 82X Data Release 3 catalog.

\subsection{Spectroscopic Follow-up Campaigns of Stripe 82X Sources}\label{new_specz}

Since 2012, we have been undertaking ground-based observing campaigns to increase the spectroscopic completeness of the survey using optical telescopes and to target obscured AGN candidates using near-infrared telescopes \citep[e.g.][]{lamassa2017}. A special SDSS-IV eBOSS spectroscopic program targeted the XMM-AO13 survey area, identifying 616 X-ray sources which increased the spectroscopic completeness of this portion of the survey to 63\% \citep{lamassa2019}. Over the past several years, we have focused our follow-up optical spectroscopic campaigns on the remaining {\it XMM}-AO13 sources and {\it XMM}-AO10 sources due to the homogeneous X-ray flux and areal coverage in this portion of the Stripe 82X survey. We used DoubleSpec on the 5-meter Palomar telescope to target optical counterparts brighter than $r = 22$ \citep[AB;][]{oke1982,rahmer2012} and the Low Resolution Imaging Spectrometer \citep[LRIS;][]{oke1995,rockosi2010} and DEep Imaging Multi-Object Spectrograph \citep[DEIMOS;][]{faber2003} on the 10-meter Keck telescopes to target optical sources fainter than $r = 22$. In the 2022 autumn observing run, we reached the milestone of observing every optical counterpart to an X-ray source from {\it XMM}-AO10 and {\it XMM}-AO13 brighter than $r = 22$.\footnote{For sources between $18 < r < 22$, we required an offset star brighter than $r < 18$ and within $<$90$^{\prime\prime}$ in RA or declination of the source to acquire the target with DBSP. Twenty-six targets from {\it XMM}-AO10 and {\it XMM}-AO13 at $r < 22$ did not have a bright star nearby to allow for blind offsetting and were thus not observed.}

This catalog release includes 343 new spectroscopic redshifts and classifications of Stripe 82X sources from both our optical and near-infrared campaigns that have not been previously published. As summarized in Table \ref{s82x_summary}, we have spectroscopic redshifts for 3457 of the 6181 X-ray sources in Stripe 82X, or 56\% of the sample. Of these sources, 3211 (93\%) are X-ray AGN ($L_{\rm X} > 10^{42}$ erg s$^{-1}$), 94 (3\%) are galaxies ($L_{\rm X} < 10 ^{42}$ erg s$^{-1}$)\footnote{Given the X-ray luminosity limit used to define a source as an AGN, objects classified as ``galaxies'' in this context can include low luminosity AGN or obscured AGN.}, and 142 (4\%) are stars. Of the X-ray AGN, 20\%  lack broad emission lines and are classified as Type 2 AGN. There are 6 X-ray sources that are optically classified as AGN (i.e., they have broad lines in their spectra) but X-ray luminosities below the L$_{\rm X}$ = 10$^{42}$ erg s$^{-1}$ AGN definition threshold. These sources are listed as ``Optical, X-ray weak AGN'' in Table \ref{s82x_summary}. Eighteen objects with redshifts did not have spectroscopic classifications in the archival databases we queried.

When considering the {\it XMM}-AO10 and {\it XMM}-AO13 sectors of the Stripe 82X survey, 3613 X-ray sources are detected. Of these sources, 2194 are detected in the optical and are brighter than $r = 22$ (AB). We have spectroscopic redshifts and classifications for 1969 of these sources, or 90\% of this subsample. Of these, 1836 are X-ray AGN (93\%), 50 are galaxies (3\%), 76 are stars (4\%), and 5 are optical, X-ray weak AGN ($<0.3$\%).

\subsection{Accuracy of Stripe 82X Photometric Redshifts}

\begin{figure}
  \includegraphics[scale=0.5]{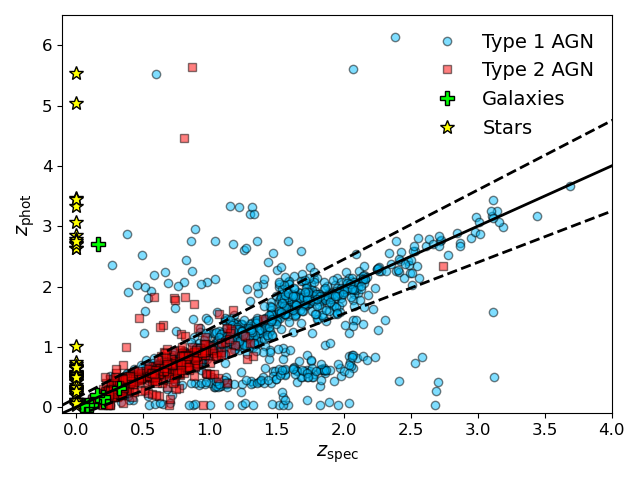}
  \caption{\label{photz_v_specz} Stripe 82X photometric redshifts ($z_{\rm phot}$) compared with spectroscopic redshifts ($z_{\rm spec}$) for sources with spectroscopic redshifts obtained after the publication of the Stripe 82X DR2 photometric redshift catalog \citep{ananna2017}. Subclasses of sources, identified by spectroscopic classification, are plotted separately as indicated by the legend. The solid line is the one-to-one relation while the dashed lines delineate the $|(z_{\rm spec}- z_{\rm phot})|/(1 + z_{\rm spec}) > 0.15$ boundaries used to define significant outliers. Overall, the accuracy of the photometric redshifts (defined by the normalized median absolute deviation) is 0.08 with an outlier fraction of 24\%. }
\end{figure}

In Figure \ref{photz_v_specz},  we test the accuracy of the Stripe 82X photometric redshifts ($z_{\rm phot}$) by comparing them with the 1589 spectroscopic redshifts ($z_{\rm spec}$) obtained following the publication of the photometric redshifts in the Stripe 82X DR2 catalog \citep{ananna2017}. The new spectroscopic redshifts are garnered from SDSS data releases that occurred after the publication of the Stripe 82X DR2 catalog, the special SDSS eBOSS program to target Stripe 82X \citep{lamassa2019}, and from our ground-based following up campaigns with Palomar and Keck (see Appendix). Similar to the procedure outlined in \citet{ananna2017}, we quantify the accuracy of the photometric redshifts using the normalized median absolute deviation ($\sigma_{\rm nmad}$):
\begin{equation}
  \sigma_{\rm nmad} = 1.48 \times {\rm median}\left(\frac{|z_{\rm spec} - z_{\rm phot}|}{1 + z_{\rm spec}}\right),
 \end{equation}
and outlier fraction by identifying the fraction of sources that exceed the following threshold:
 \begin{equation}
   \frac{|z_{\rm spec} - z_{\rm phot}|}{1 + z_{\rm spec}} > 0.15
 \end{equation}

 The overall accuracy for this sample is 0.08 with an outlier fraction of 24\%. This accuracy and outlier fraction are a bit worse than the the sample used to train and test the photometric redshifts in \citet{ananna2017}, where $\sigma_{\rm nmad}$ = 0.06 and the outlier fraction is 14\%. This trend could be due to the sources for which we have new spectroscopic redshifts being fainter (median $r = 21.3$, AB) than those used to train and test the photometric redshift sample (median $r=19.9$, AB).

 For the Type 1 AGN, there is a sequence of incorrect photometric redshifts from $0.3 \lesssim z_{\rm phot} \lesssim 0.85$ and $0.8 \lesssim z_{\rm spec} \lesssim 2$ which can be explained by features that are misidentified during the SED template fitting procedure, e.g., H$\beta$ emission at higher redshift ($0.8 < z < 1.5$) was misconstrued as H$\alpha$ emission at lower redshift ($0.3 < z < 0.85$), and does not indicate a systematic error due to the quality of the photometric data points. We also find that some high-redshift AGN candidates based on their photometric redshifts have stellar spectra.

\begin{deluxetable*}{lll}
\tablecaption{\label{s82x_summary}Stripe 82X Demographics\tablenotemark{a}}
\tablehead{\colhead{Category} & \colhead{Full Survey} & \colhead{{\it XMM}-AO10 \& {\it XMM}-AO13} \\
  & & \colhead{($r < 22$, AB)}}
\startdata
X-ray Sources                  & 6181 & 2194 \\
Spectroscopic Redshifts\tablenotemark{b}  & 3457 & 1969 \\
Spectroscopic Completeness     & 56\% & 90\% \\
X-ray AGN\tablenotemark{c}     & 3211 & 1836 \\
\hspace{5mm} Type 1 AGN & 2569 & 1490\\ 
\hspace{5mm} Type 2 AGN        & 628 &  342 \\
Optical, X-ray weak AGN\tablenotemark{d} & 6 & 5 \\
Galaxies                       & 94  & 50   \\
Stars                          & 142  & 76   \\
\hline
Area                           & 31.3 deg$^2$ & 20.2 deg$^2$ \\
\enddata
\tablenotetext{a}{Classifications of sources are for those that have spectroscopic redshifts.}
\tablenotetext{b}{Eighteen objects with archival spectroscopic redshifts from the pre-existing 6dF \citep{jones2004,jones2009} and WiggleZ \citep{drinkwater2010} catalogs and the catalogs of \citet{jiang2006} and \citet{ross2012} did not provide optical spectroscopic classifications; 6 of these unclassified objects are in the {\it XMM}-AO10 and {\it XMM}-AO13 portions of the survey.}
\tablenotetext{c}{AGN are defined as having a $k$-corrected 2-10 keV luminosity ($L_{\rm X}$) value greater than 10$^{42}$ erg s$^{-1}$. $L_{\rm X}$ values are not corrected for absorption and are based on converting the X-ray count rate to flux assuming a power law model. Objects considered ``Galaxies'' in this context may include obscured AGN or AGN with low X-ray luminosities.}
\tablenotetext{d}{Sources with broad lines in their optical spectra that are thus classified as ``QSO'' in the catalog but have X-ray luminosities below the canonical AGN threshold of 10$^{42}$ erg s$^{-1}$.}
\end{deluxetable*}

\section{Obscured AGN Demographics from Stripe 82X {\it XMM}-AO10 and {\it XMM}-AO13 at $r < 22$}\label{obsc_agn_sec}

\subsection{Defining Obscured AGN}

Due to the homogeneous X-ray coverage and high level of spectroscopic completeness at $r < 22$ (AB) in the {\it XMM}-AO10 and {\it XMM}-A013 portions of Stripe 82X, we explore the obscured AGN demographics within this subset of the survey. We focus on the 1836 X-ray AGN as the parent sample and consider three flavors of obscured AGN:
\begin{enumerate}
\item Optically obscured/Type 2 AGN (i.e., those that do no have broad lines in their spectra, which includes galaxies with narrow emission lines and those without emission lines, i.e., the so-call X-ray Bright Optically Normal Galaxies \citep[XBONGs;][]{comastri2002, hornschemeier2005});
\item Reddened AGN ($R-K > 4$, Vega), including both Type 1 and Type 2 AGN, though we caution that red $R-K$ colors for low-redshift AGN could be due to starlight rather than dust extinction \citep[see, e.g.,][]{glikman2022} and distinguishing between these scenarios is beyond the scope of this paper;
\item X-ray obscured AGN ($N_{\rm H} > 10^{22}$ cm$^{-2}$).
\end{enumerate}

We note that 6 X-ray AGN have spectroscopic redshifts but not classifications in the archival catalogs we queried, so we are we are therefore unable to classify them as optically unobscured or optically obscured. We thus can classify 1830 AGN as Type 1 or Type 2 based on the presence or absence of broad lines in their optical spectra in Table \ref{obsc_agn}.

We follow the prescription in \citet{lamassa2016b} and \citet{lamassa2017} to convert the SDSS $r^{\prime}$ (AB) magnitude to the Bessell $R$ magnitude (Vega), using an $r-i$ color correction, to allow a more direct comparison with $R-K$-selected reddened AGN from the literature \citep[e.g.,][]{glikman2007,glikman2013,brusa2010}. As discussed in \citet{lamassa2016b}, the SDSS magnitudes are measured assuming a PSF model for point sources, or with an exponential profile or de Vaucoulers profile for extended sources, while the near-infrared magnitudes are measured through a larger 2.8$^{\prime\prime}$ or 5.6$^{\prime\prime}$ aperture for point and extended sources, respectively. Different portions of the host galaxy can thus be contributing to the optical and infrared magnitude measurements. However, as illustrated in \citet{lamassa2016b}, there are not signficant differences between the SDSS model magnitudes and a larger aperture ``auto'' magnitude measured via Source Extractor \citep{bertin1996} and published in the SDSS coadded catalog of \citep{jiang2014}. The convention followed here is also consistent with precedent in the literature \citep[e.g.,][]{urrutia2009,glikman2013}.

There are two near-infrared surveys that cover Stripe 82: the VISTA Hemisphere Survey \citep[VHS;][]{mcmahon2013} and the UKIRT Infrared Deep Sky Survey \citep[UKIDSS;][]{lawrence2007}. The VHS survey is slightly deeper than UKIDSS, with a 5$\sigma$ depth of $K=20.3$ (AB) compared with a 5$\sigma$ depth of $K=20.1$, respectively, and the VHS photometry has smaller errors than UKIDSS \citep[see Figure 7 in][]{ananna2017}. Similar to \citet{ananna2017}, we choose the VHS $K$-band photometry if available, but otherwise use the UKIDSS photometry. Since the magnitudes in the Stripe 82X catalog are reported in the AB system, we convert the $K$-band magnitude to Vega using $K_{\rm Vega} = K_{\rm AB} - 1.9$. There are 1645 X-ray AGN that have available optical and infrared photometry to identify reddened AGN.

We use the X-ray spectral fitting catalog from \citet{peca2023} for the $N_{\rm H}$ measurements to identify which AGN are X-ray obscured. For the spectral fitting to be performed, at least 20 net counts were required, though the minimum number of counts needed for a spectral fit varies as a function of AGN redshift and X-ray spectal model complexity \citep[see][for details]{peca2023}. 1168 AGN have column density measurements from X-ray spectral fitting and can thus be classified as X-ray obscured ($N_{\rm H} > 10^{22}$ cm$^{-2}$) or unboscured.

In Table \ref{obsc_agn}, we list the number of AGN from the ``parent'' sample for each subclass, i.e., the number of sources where we have information to classify an AGN as ``obscured'' according to the metric considered, and the number of obscured AGN in each category. We find that 417 AGN (23\%) are considered obscured by at least one of these metrics. The largest category of obscured AGN are Type 2 AGN, making up 19\% of the population.

We note that the X-ray obscured AGN fraction reported here (10\%) is likely a lower limit since heavily obscured AGN will have X-ray spectra of low quality or too few counts to permit spectral modeling to measure an obscuring column density. Indeed, after correcting for Stripe 82X survey biases, \citet{peca2023} find an {\it intrinsic} X-ray obscured AGN fraction of $\sim$57\% for X-ray luminosities above 10$^{43}$ erg s$^{-1}$. Our analysis is complementary to that of \citet{peca2023} as they consider only X-ray obscured AGN while we include Type 2 and reddened AGN in this analysis. Additionally, we focus on just the {\it XMM}-AO10 and {\it XMM}-AO13 portions of the Stripe 82X survey at $r < 22$ (AB) while \citet{peca2023} consider the full Stripe 82X survey.

\subsection{(Dis)Agreement Among Obscured AGN Classifications}

\begin{figure*}
  \includegraphics[scale=0.58]{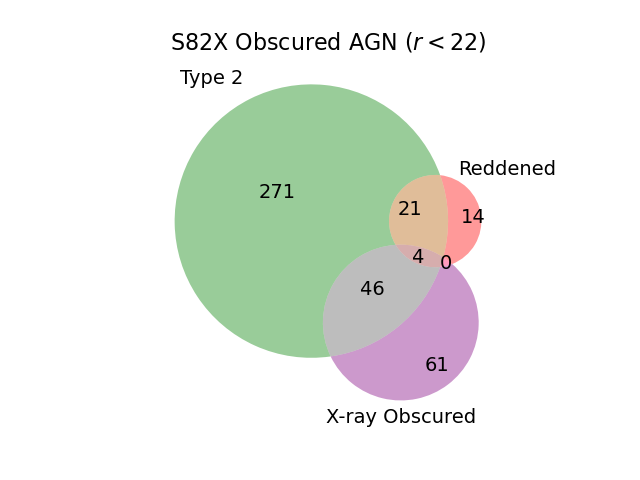}
  \includegraphics[scale=0.58]{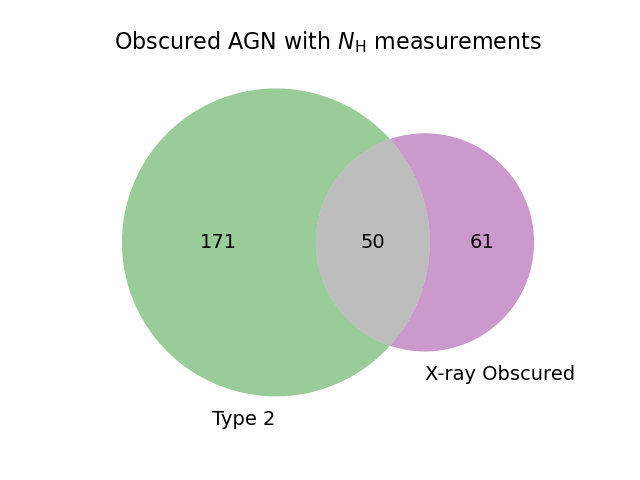}
  \caption{\label{obsc_venn} ({\it Left}:) Venn diagram showing classes of obscured AGN for all Stripe 82X AGN at $r < 22$ with spectroscopic redshifts for the {\it XMM}-AO10 and {\it XMM}-AO13 portions of the survey. We distinguish among the following categories of obscured AGN: Type 2 AGN (i.e., no broad lines in optical spectra), reddened ($R-K > 4$, Vega), and X-ray obscured ($N_{\rm H} > 10^{22}$ cm$^{-2}$). ({\it Right}): Subset of obscured AGN that have both spectroscopic classifications and measured $N_{\rm H}$ values from X-ray spectral fitting; 121 of the Type 2 AGN shown in the left diagram do not have $N_{\rm H}$ measurements and are thus omitted from this figure. Of the 282 obscured AGN in this subset, only 18\% are both Type 2 and X-ray obscured.}
\end{figure*}

In Figure \ref{obsc_venn} (left), we show a Venn diagram of these obscured AGN classifications, highlighting the population that is unqiue to one metric and where there is agreement among the definitions. We find that more than half of the X-ray obscured AGN are optically {\it unobscured} suggesting a distinction between the gas that attenuates X-ray photons and the more distant dust that obscures the broad line region, perhaps related to the radiation environment around the black hole \citep[e.g.,][]{merloni2014}. Evidence of such high column density clouds within the broad line region dust sublimation zone have been observed in so called X-ray ``changing-look'' AGN \citep[e.g.,][]{mckernan1998, risaliti2009, marinucci2016}.

We find that there are no reddened Type 1 AGN that are also X-ray obscured in the {\it XMM}-AO10 and {\it XMM}-AO13 segments of the Stripe 82X survey. We refer the reader to \citet{glikman2018} for a discussion on {\it WISE}-selected AGN from the full 270 deg$^2$ Stripe 82 survey area, 13 of which have archival {\it XMM-Newton} and {\it Chandra} detections, as well as {\it Chandra} GTO data: based on the X-ray hardness ratios,\footnote{The X-ray hardness ratio is $(H-S)/(H+S)$, where $H$ ($S$) refers to the counts in the hard (soft) X-ray band.} 3 out of 4 reddened Type 1 AGN in that sample of 13 are X-ray obscured.

When we limit this comparison to just the AGN that have both spectroscopic classifications and $N_{\rm H}$ measurements from X-ray spectral fitting (Figure \ref{obsc_venn}, right), we see that only 18\% of AGN are both obscured optically and in X-rays. This is a much lower fraction than reported in previous X-ray surveys. Samples of hard X-ray selected AGN (E $>$ 10 keV) in the local Universe from INTEGRAL \citep[predominantly $z < 0.1$;][]{bird2010} and Swift-BAT \citep[mostly $z < 0.4$;][]{baumgartner2013,oh2018} find a 12\% disrepancy and $\sim$2\% discrepancy, respectively between optically and X-ray obscured AGN \citep{malizia2012,koss2017}. From the moderate-area, moderate-depth {\it XMM}-COSMOS survey \citep{hasinger2007,cappelluti2009}, spanning a similar redshift range as Stripe 82X ($0.3 < z < 3.5$) though sampling the moderate X-ray luminosity AGN population, \citet{merloni2014} find a 30\% disagreement between optical and X-ray obscured AGN classifications. The most direct comparison to Stripe 82X is from the {\it XMM}-XXL-N field which covers a comparable area ($\sim$25 deg$^2$) at a comparable depth \citep{pierre2016,menzel2016}. Compared with the other X-ray surveys, there is a larger discrepancy between X-ray and optically obscured AGN \citep{liu2016}, but this fraction of 40\% is still well below the $\sim$80\% fraction we find in Stripe 82X. We note that the dividing column density between X-ray obscured and unobscured varies from $N_{\rm H} = 10^{21.5}$ cm$^{-2}$ \citep{merloni2014,liu2016} to $N_{\rm H} = 10^{21.9}$ cm$^{-2}$ \citep{koss2017}. However, when we experiment with a threshold of $N_{\rm H} \geq 10^{21}$ cm$^{-2}$ to classify an AGN as obscured, the fraction of overlapping Type 2 and X-ray obscured AGN only increases to $\sim$23\% meaning that the larger disagreement in the overlap of obscured AGN between Stripe 82X and other surveys is not driven by the obscured AGN column density threshold. We also considered AGN whose upper limit on the fitted column density exceeds 10$^{22}$ cm$^{-2}$: while this more expansive definition of X-ray obscured AGN increased the overall number to 596, the overlap between Type 2 and X-ray obscured AGN was only 19\%.

\begin{deluxetable*}{llll}
\tablecaption{\label{obsc_agn}Stripe 82 {\it XMM}-AO10 \& {\it XMM}-AO13 Obscured AGN Summary ($r < 22$, AB)}
\tablehead{\colhead{Category} & \colhead{Parent sample\tablenotemark{a}} & \colhead{Number of Obscured AGN} 
& \colhead{Obscured AGN Fraction\tablenotemark{b}}}
\startdata
Optically Obscured & 1830 & 342 & 19\% \\
(Type 2 AGN) \\
Reddened AGN  & 1645 & 39 & 2.4\% \\
($R - K > 4$) \\
X-ray Obscured & 1168 & 111 & 10\% \\
($N_{\rm H} > 10^{22}$ cm$^{-2}$) \\
\enddata
\tablenotetext{a}{The number of X-ray AGN with optical spectroscopic classifications; $i$-band measurements (for calculating the $r-i$ color correction to derive $R$) and $K$-band measurements; and with X-ray spectral fitting results to measure $N_{\rm H}$, to categorize an AGN as optically obscured, reddened, or X-ray obscured, respectively. The parent sample for the reddend AGN subclass is lower than that of the optically obscured subclass largely due to $K$-band non-detections in the former sample.}
\tablenotetext{b}{The fraction of obscured AGN is calculated with respect to the parent sample for that class, that is, the number of X-ray AGN where we have the information to classify an AGN as obscured according to the metric considered.}
\end{deluxetable*}

\subsection{X-ray Luminosity - Redshift Distribution of Obscured AGN}

\begin{figure*}
  \includegraphics[scale=0.45]{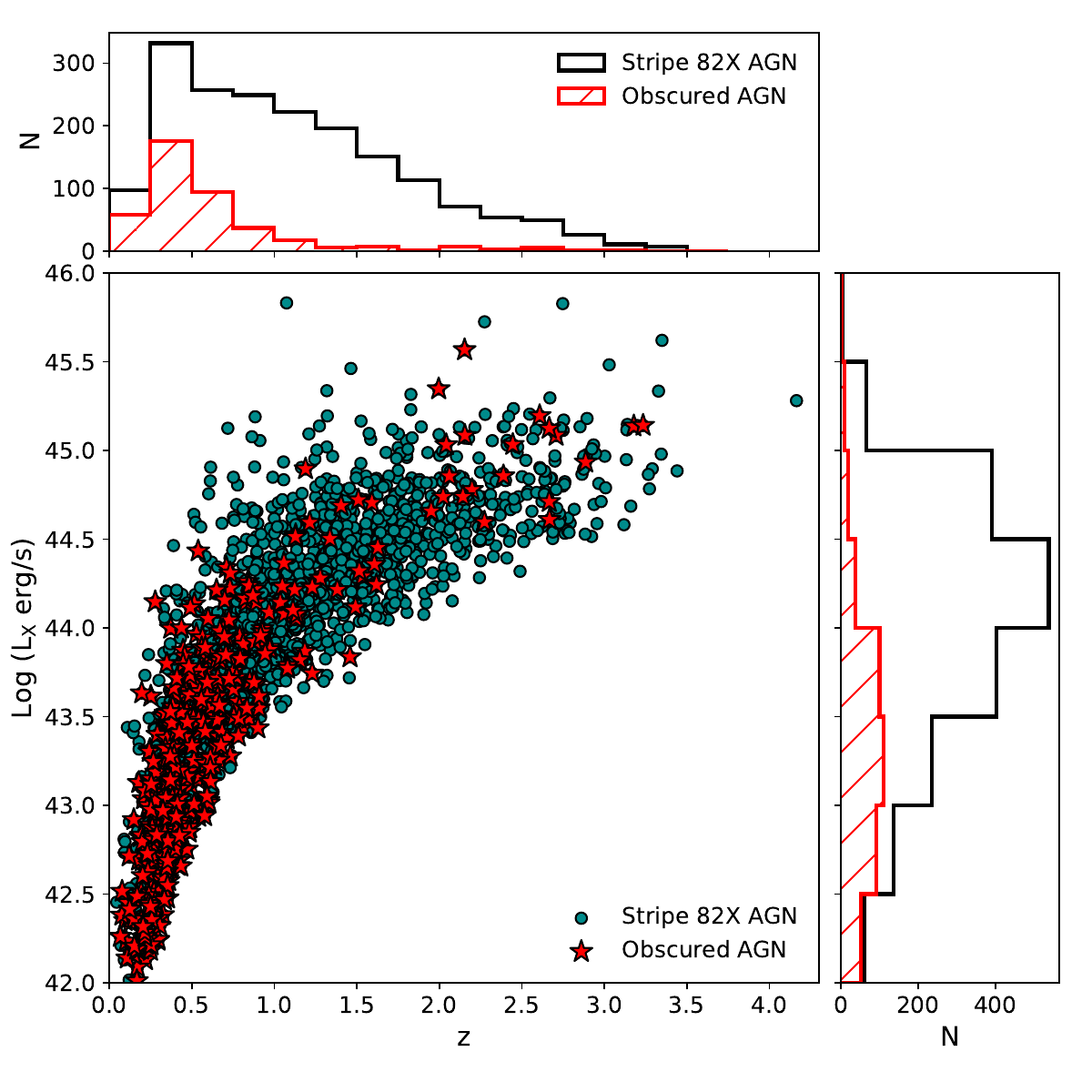}~
  \includegraphics[scale=0.45]{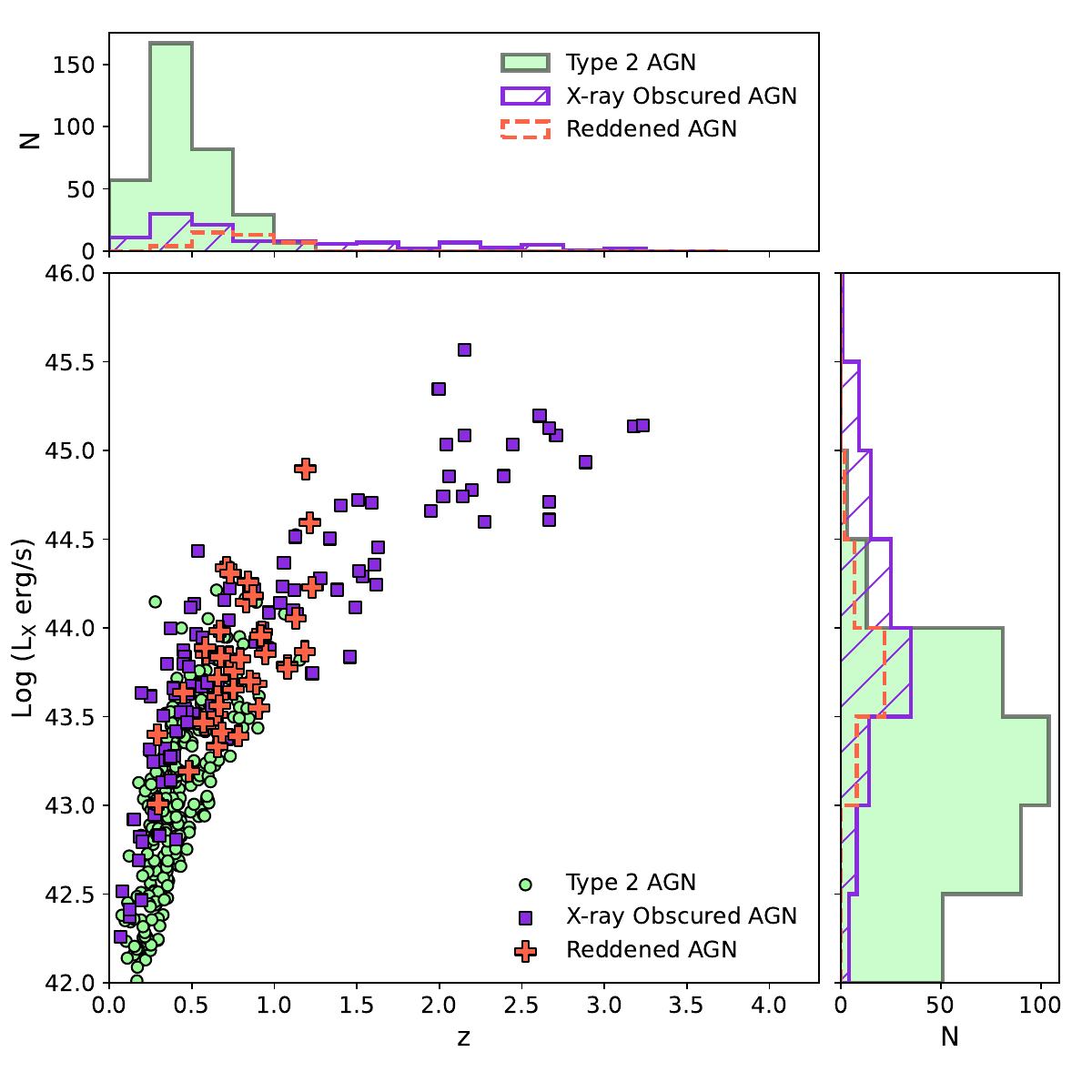}
  \includegraphics[scale=0.45]{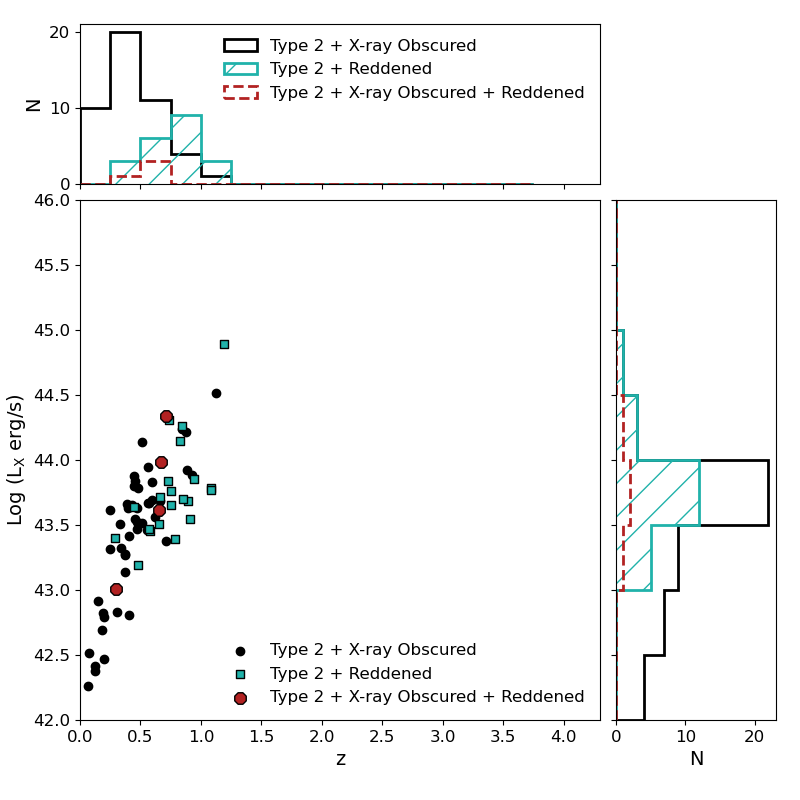}
  \caption{\label{lx_v_z} Rest-frame, non-absorption corrected 2-10 keV X-ray luminosity as a function of redshift for all spectroscopically identified AGN from Stripe 82 {\it XMM}-AO10 and {\it XMM}-AO13 at $r < 22$. At these X-ray and optical flux limits, the Stripe 82X survey is 90\% spectroscopically complete. ({\it Left}): Since Stripe 82X is a shallow, wide-area survey, it largely samples the high X-ray luminosity AGN parameter space. The obscured AGN (Type 2 AGN, reddened AGN, and X-ray obscured AGN; red stars) are predominantly at lower X-ray luminosities and redshifts and make up the majority of low X-ray luminosity ($L_{\rm X} <10^{43}$ erg s$^{-1}$) Stripe 82X AGN (see also Table \ref{obsc_frac}). ({\it Right}): Of the obscured AGN population, the Type 2 AGN (green circles) are at the lowest X-ray luminosities and redshifts. The X-ray obscured AGN (where $N_{\rm H} > 10^{22}$ cm$^{-2}$; purple squares) extend to high X-ray luminosities ($L_{\rm X} > 10^{44.5}$ erg s$^{-1}$) and high redshifts ($z > 1$). The reddened AGN (red crosses) are found at moderate-to-high X-ray luminosities, overlapping the tail of the Type 2 AGN $L-z$ distribution. There is some overlap in these sub-populations of obscured AGN (see Figure \ref{obsc_venn}) ({\it Bottom}): X-ray luminosity-redshift distribution for AGN that are classified as obscured based on multiple metrics (as indicated in the legend). The Type 2 + X-ray obscured AGN span a range of redshifts and X-ray luminosities, while the AGN that are reddened and also Type 2 and/or X-ray obscured are at higher X-ray luminosities ($L_{\rm X} = 10^{43}$ erg s$^{-1}$) and mostly above $z > 0.5$, but they do not extend to as high redshift or X-ray luminosity as the AGN that are only X-ray obscured.}
\end{figure*}

We explore the X-ray luminosity-redshift parameter space inhabited by the spectroscopically identified Stripe 82X AGN at $r<22$ from {\it XMM}-AO10 and {\it XMM}-AO13 in Figure \ref{lx_v_z}. We use the $k$-corrected, non-absorption corrected 2-10 keV luminosity ($L_{\rm X}$) as a proxy of the AGN luminosity. We note that the intrinsic X-ray luminosity can be higher, but we only have $N_{\rm H}$ measurements and thereby intrinsic X-ray luminosities for a subset of the sample so we focus on the observed luminosity to not limit this analysis.

By design, Stripe 82X samples the high X-ray luminosity AGN population. The top left panel of Figure \ref{lx_v_z} shows that, consistent with previous surveys and intrinsic X-ray luminosity functions that probe AGN beyond the local Universe \citep[e.g.,][]{merloni2014,ueda2014,buchner2015,liu2017}, the obscured AGN fraction dominates the population at lower X-ray luminosities and then decreases as X-ray luminosity increases (see Table \ref{obsc_frac} for a summary). These results are consistent with the ``receding torus'' model where increased radiation from higher luminosity AGN reduces the torus covering factor \citep{lawrence1991,hoenig2007}.

However, about a quarter of obscured AGN emit at the highest X-ray luminosities ($L_{\rm X} > 10^{44}$ erg s$^{-1}$). Furthermore, as X-ray selection is biased against the Compton-thick population, heavily obscured AGN at higher luminosities can be altogether missing from Stripe 82X, neccesitating AGN samples selected at other wavelengths for a more comprehensive determination of the evolution of the intrinsic obscured AGN fraction with luminosity \citep[e.g.,][]{lusso2013,mateos2017}.

In the right hand panel of Figure \ref{lx_v_z}, we consider the obscured AGN subpopulations separately.  We find that the Type 2 AGN are predominantly at $z < 1$ and X-ray luminosities below $L_{\rm X} < 10^{44}$ erg s$^{-1}$, though there is a tail of the distribution at higher X-ray luminosities and up to redshifts of $z \sim 1.5$. The reddened AGN lie at redshifts $0.25 < z < 1.25$ and are at moderate to high X-ray luminosities ($L_{\rm X} \geq 10^{43.5}$ erg s$^{-1}$), though the number of Type 2 AGN at these  X-ray luminosities is comparable to the number of reddened AGN. The most X-ray luminous ($L_{\rm X} > 10^{44.5}$ erg s$^{-1}$) obscured AGN are predominantly X-ray obscured: these AGN are Type 1 and are mostly not reddened, but based on X-ray spectral modeling, are attenuated by column densities $N_{\rm H} > 10^{22}$ cm$^{-2}$.

In the bottom panel of Figure \ref{lx_v_z}, we explore the X-ray luminosity-redshift space of AGN that are classified as obscured according to multiple metrics. The Type 2 + X-ray obscured AGN span the largest range in redshift ($0 < z < 1.25$) and X-ray luminosity ($10^{42}$ erg s$^{-1} < L_{\rm X} < 10^{44.5}$ erg s$^{-1}$) while the reddened AGN + Type 2 and reddened AGN + Type 2 + X-ray obscured are at higher redshift ($z > 0.25$) and X-ray luminosity ($L_{\rm X} > 10^{43}$ erg s$^{-1}$). Still, we see that the AGN that are only X-ray obscured (top right panel of Figure \ref{lx_v_z}) extend to higher redshifts ($z > 1.25$) and X-ray luminosities ($L_{\rm X} > 10^{44.5}$ erg s$^{-1}$) than those that are also reddened.

These results are consistent with the trends reported in \citet{merloni2014} and summarized in \citet{hickox2018} that Type 2 AGN tend to be at lower X-ray luminosities while the fraction of X-ray obscured and optically unobscured AGN increases with X-ray luminosity. These results are also consistent with those reported by \citet{glikman2018} who found that, using a sample of mid-infrared color selected AGN, the fraction of reddened Type 1 AGN increases with X-ray luminosity while the opposite trend is seen for Type 2 AGN.

\begin{deluxetable}{lll}
\tablecaption{\label{obsc_frac}Observed Obscured AGN Fraction\tablenotemark{a}}
\tablehead{\colhead{$L_{\rm X}$ Bin\tablenotemark{b}} & \colhead{Obscured AGN \#} & \colhead{Obscured Fraction} \\
  \colhead{Log (erg s$^{-1}$)} 
}
\startdata
42.0 - 42.5 & 51 & 84\% \\
42.5 - 43.0 & 91 & 66\% \\
43.0 - 43.5 & 110 & 47\% \\
43.5 - 44.0 & 100 & 25\% \\
44.0 - 44.5 & 37 & 6.9\% \\
44.5 - 45.0 & 18 & 4.6\% \\
$>$45.0     & 10 & 14\% \\
\enddata
\tablenotetext{a}{The reported AGN demographics is for the Stripe 82 {\it XMM}-AO10 and {\it XMM}-AO13 sectors of the survey at $r < 22$ which is 90\% spectroscopically complete. Here we consider an AGN obscured if it is either a Type 2 AGN (no broad lines in the AGN spectrum), reddened ($R- K > 4$, Vega), or X-ray obscured with a measured column density of $N_{\rm H} > 10^{22}$ cm$^{-2}$ from X-ray spectral fitting \citep{peca2023}. The percentages are calculated for a parent sample that includes all Stripe 82X AGN ($L_{\rm X} > 10^{42}$ erg s$^{-1}$) with spectroscopic redshifts from {\it XMM}-AO10 and {\it XMM}-AO13.}
\tablenotetext{b}{X-ray luminosities are not corrected for absorption.}
\end{deluxetable}

\section{Measuring Black Hole Masses for Type 1 AGN}\label{bhmass_meas}

We measure the black hole masses ($M_{\rm BH}$) of Type 1 AGN in Stripe 82X from SDSS single epoch spectroscopy using virial relations that were calibrated on results from reverberation mapping (RM) campaigns \citep[see][for a review]{peterson2014}. These relations rely on measuring the continuum luminosity and width of broad emission lines in Type 1 AGN and take the form of:
\begin{eqnarray}\label{se_formula}
  \rm{Log}\left(\frac{M_{\rm BH,vir}}{M_{\rm \sun}}\right) = a + b\times \rm{Log}\left(\frac{L}{10^{d}\,\rm{erg}\rm{s}\,^{-1}}\right)+ \nonumber\\
    c\times \rm{Log}\left(\frac{\rm{FWHM}}{{\rm km}\,{\rm s}^{-1}}\right),
\end{eqnarray}
where $L$ refers to the wavelength at which the AGN continuum luminosity is driving the line emission, FWHM is the measured full width half maximum of the broad emission line, $a$, $b$, and $c$ are fitted constants from extrapolating RM results to scaling relationships, and $d$ is the normalization on the continuum luminosity in log space. Table \ref{se_parameters} summarizes the scaling relations we used in this work which are mostly from recalibrations to single epoch formulas from the latest release of the SDSS Reverberation Mapping project, which extends to higher redshifts and bolometric luminosities compared with previous RM campaigns \citep{shen2024}. The uncertainties in black hole mass measurements is on the order of 0.45 dex for $M_{\rm BH}$ values derived from the H$\beta$ and \ion{Mg}{2} lines and $\sim$0.58 dex for \ion{C}{4}-derived black hole masses.  More information about each of the single epoch relations is given in Section \ref{se_discussion} to motivate the preference we gave in selecting the reference $M_{\rm BH}$ value in the catalog.

\begin{table*}
  \caption{Single epoch $M_{\rm BH}$ formula parameters$^{a}$}
  \label{se_parameters}
  \begin{tabular}{lllllll}
   \hline
   Line & $L^{b}$ & a & b & c & d& Reference \\
   \hline  
  H$\beta$ & $L_{\rm 5100\angstrom}$ & 0.85 & 0.50 & 2.0 & 44 & \citet{shen2024} \\
  H$\alpha$ & $L_{\rm 5100\angstrom}$ & 0.807 & 0.519 & 2.06 & 44 & \citet{greene2010}\\
  \ion{Mg}{2} & $L_{\rm 3000\angstrom}$ & -2.05 & 0.60 & 3.0 & 45 & \citet{shen2024} \\
  \ion{C}{4} & $L_{\rm 1350\angstrom}$ & 1.40 & 0.50 & 2.0 & 45 & \citet{shen2024} \\
  \hline
  \multicolumn{7}{l}{$^a$Parameters used in Equation \ref{se_formula} to calculate black hole masses.}\\
    \multicolumn{7}{l}{$^b$The wavelength at which the AGN continuum luminosity that powers}\\
    \multicolumn{7}{l}{the line emission is calculated.}
  \end{tabular}
\end{table*}

\subsection{Virial Mass Scaling Relations for Single Epoch Spectroscopy}\label{se_discussion}
Reverberation mapping campaigns have historically been restricted to nearby ($z < 0.3$) AGN to allow for high signal-to-noise spectra (S/N $\sim$ 100) to be observed over the span of months to years to measure variability. As a result, most AGN with black hole masses measured from RM campaigns are based on the response of the H$\beta$ line to the variability in the continuum \citep[e.g.,][]{peterson2002,denney2010,grier2012,grier2017,bentz2015}. As H$\beta$ is the most common line used for RM black hole masses, this single epoch spectroscopy scaling relation from \citet{shen2024} is the most reliable of the four we consider in this analysis.

The next most reliable single epoch spectroscopy scaling relation is that based on the ultraviolet \ion{Mg}{2} line \citep{shen2024}. At modest redshifts ($z>$0.9), H$\beta$ is no longer accessible in observed-frame optical spectra, making ultraviolet emission lines a useful diagnostic for measuring black hole masses at this redshift regime from spectra in large ground based optical surveys like SDSS. \citet{mclure2002} pointed out that \ion{Mg}{2} is a fair substitute for the H$\beta$ line as a virial black hole mass estimator due to their similar ionization potentials, indicating that the gas responsible for these emission lines lie at the same radius in the broad line region. 

The H$\alpha$ scaling relation from \citet{greene2010} is based on the radius-luminosity calibration in \citet{bentz2009} which was defined using the H$\beta$ line. To derive black hole mass from H$\alpha$, \citet{greene2010} converted the FWHM of H$\beta$ to H$\alpha$ using the formulism in \citet{greene2005}. Most of the Type 1 AGN in Stripe 82X are at too high a redshift for H$\alpha$ to be observed, and we defer to the H$\beta$ scaling relation when possible since this calibration is based directly on RM black hole measurements.

Finally, the ultraviolet \ion{C}{4} line can be used as a black hole estimator for higher redshift AGN ($z>$1.5) observed by SDSS. The \ion{C}{4} line mass estimator is controversial since asymmetries and blueshifts in the line profile \citep{marziani1996} and disagreements between \ion{C}{4}-based masses and those calculated from H$\beta$, H$\alpha$ and MgII \citep{netzer2007, shen2008, shen2012, trakhtenbrot2012} indicate that the gas producing \ion{C}{4} has significant contributions from non-virialized motion \citep{baskin2005, shen2008}, likely due to outflows for AGN exceeding $\sim$10\% Eddington \citep{temple2023}. However, monitoring campaigns of high redshift AGN show that the \ion{C}{4} line does reverberate in response to the AGN continuum variability \citep{kaspi2007,derosa2015,grier2019}, allowing \ion{C}{4}-based black hole masses to be calculated via RM campaigns. We use the \ion{C}{4} viral mass estimator derived in \citet{shen2024}, and note that a comparison between \ion{C}{4} RM black hole masses and those from single epoch spectroscopy show that the latter values are systematically larger than the former \citep{lira2018}, perhaps due to contributions from non-virial motion to the \ion{C}{4} line profile \citep{denney2012,lira2018}. Indeed, as demonstrated in \citet{shen2024}, even with the most complete \ion{C}{4} reverberation mapping black hole mass results to date, systematic biases in the \ion{C}{4}-derived $M_{\rm BH}$ estimate remain, resulting in the widest scatter between the single-epoch measurements and RM measurements ($\sim$0.58 dex).

\subsection{Spectral Decomposition of Stripe 82X Type 1 AGN}
To derive the spectroscopic parameters needed to calculate $M_{\rm BH}$, we fitted the SDSS spectra of the 2235 Stripe 82X Type 1 AGN with \textsc{PyQSOFit} \citep{guo2018}. This fitting package is the \textsc{python} version of the \textsc{IDL} codes used to calculate black hole masses in previous large SDSS quasar catalogs \citep[e.g.,][]{shen2011}, so it is well trained on SDSS spectra and produces results that are directly comparable to published catalogs.

To perform the spectral decomposition, \textsc{PyQSOFit} fits a continuum that consists of powerlaw emission from the accretion disk, optical \ion{Fe}{2} templates \citep{boroson1992} and ultraviolet \ion{Fe}{2} templates \citep{vestergaard2001} that produce a pseudo-continuum, a Balmer Continuum from 1000$\angstrom$-3646$\angstrom$ \citep{dietrich2002}, and a host galaxy that is fit by doing a principle component analysis (PCA) using eigenspectra derived from SDSS galaxy spectra \citep{yip2004}. We modify this routine to replace the PCA method and to instead fit galaxy spectral templates from the E-MILES library \citep{vazdekis2016}. However, in most cases the continuum was dominated by the AGN and the host galaxy was not detected, meaning that the host galaxy decomposition was not applied. The continuum is subtracted in windows around each emission line of interest, with the lines then fitted by broad and narrow components. \textsc{PyQSOFit} reports the fitted parameters on the continuum components, emission lines, and continuum luminosities ($L_{\rm 5100\angstrom}$, $L_{\rm 3000\angstrom}$, and $L_{\rm 1350\angstrom}$), with errors on fit parameters calculated via Monte Carlo resampling.

After fitting the SDSS spectra of Stripe 82X Type 1 AGN, we vetted the results. We rejected an emission line or continuum measurement if the:
\begin{itemize}
\item Fit parameter or error is non-finite (e.g., NaN);
\item Error on a fit parameter (FWHM, $\sigma$, equivalent width, central wavelength of emission line, line flux) is greater than the fitted value of that parameter;
\item FWHM is 0;
\item Continuum luminosity is 0;
\item FWHM $>$ 15,000 km/s which indicates that the Gaussian model is fitting residuals in the continuum and not the emission line.
\end{itemize}

For the line and continuum measurements that passed these checks, we then calculated black hole masses using Eq. \ref{se_formula} and the parameters in Table \ref{se_parameters}. We rejected any $M_{\rm BH}$ measurements where the error on the black hole mass exceeded the measured mass of the black hole.

We then visually inspected the fits to the remaining sources and discarded any $M_{\rm BH}$ value where the relevant emission line fit was poor. In these cases, the measured ``line'' was actually noise in the continuum or the continuum near the line was poorly measured which would bias both the continuum luminosity and the ability to accurately measure the emission line FWHM. We stress that within a spectrum, a fit can be discarded for one emission line but acceptable for another emission line.

\subsection{Black Hole Masses of Stripe 82X Type 1 AGN}
\begin{figure}
  \includegraphics[scale=0.5]{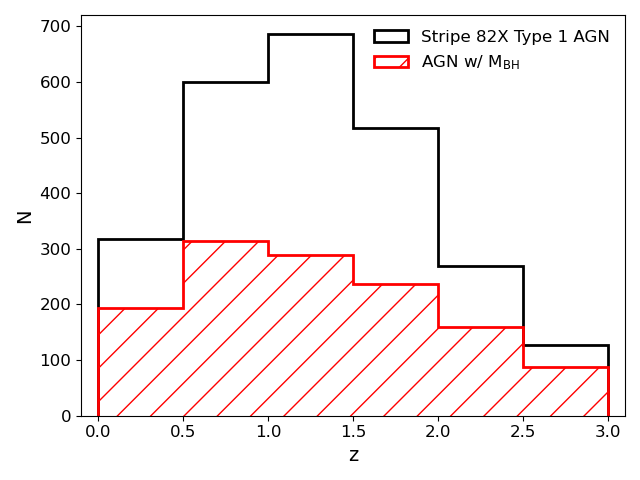}
  \caption{\label{z_dist}Redshift distribution of all Type 1 AGN in Stripe 82X with those that have black hole masses measured shown in the hatched region. The distributions are similar, indicating that our black hole mass subsample is not biased in redshift space relative to broad line AGN in Stripe 82X.}
\end{figure}

Of the 2235 SDSS spectra we fitted, we obtained an acceptable black hole measurement from at least 1 emission line for 1297 sources, representing 58\% of the Type 1 Stripe 82X AGN with SDSS spectra. The number of AGN with black hole masses measured from each emission line single epoch formula are listed in Table \ref{bhm_summary}. In our published catalog, we report the values for each of these measurements, as well as a reference M$_{\rm BH}$ value based on the ranking system described above (i.e., M$_{\rm BH}$ = 1. M$_{\rm BH,H\beta}$; 2. M$_{\rm BH,MgII}$; 3. M$_{\rm BH, H\alpha}$; 4. M$_{\rm BH,CIV}$). In Table \ref{bhm_summary}, we list the number of sources for each line estimator that are reported as M$_{\rm BH}$, as well as the redshift range and median redshift for the AGN in each subsample. We plot in Figure \ref{z_dist} the redshift distribution of all Stripe 82X Type 1 AGN, and those for which we calculated black hole masses, to illustrate that our measurements fully sample the spectroscopic redshift parameter space of Type 1 AGN in Stripe 82X.

\begin{deluxetable}{lllll}
\tablecaption{\label{bhm_summary}Black Hole Mass Tally}
\tablehead{\colhead{Line} & \colhead{Individual} & \colhead{M$_{\rm BH}$\tablenotemark{b}} & \colhead{$z$ range} & \colhead{Median $z$} \\
  & \colhead{Measurements\tablenotemark{a}}
}
\startdata
H$\beta$  & 312 & 312 & 0.03 - 1.00 & 0.52 \\
MgII      &  696 & 581 & 0.37 - 2.24 & 1.24 \\
H$\alpha$ & 113 &  38 & 0.05 - 0.42 & 0.31 \\
CIV       & 424 & 366 & 1.64 - 3.44 & 2.19 \\
\enddata
\tablenotetext{a}{This column lists the number of black hole measurements from each emission line single epoch formula where we obtained an acceptable fit to the spectrum.}
\tablenotetext{b}{Column summarizes the number of measurements from each emission line single epoch formula that are reported as our reference M$_{\rm BH}$ in the catalog.}
\end{deluxetable}

\subsection{Comparison of Emission Line Single Epoch Mass Estimators}
\begin{figure*}
  \includegraphics[scale=0.7]{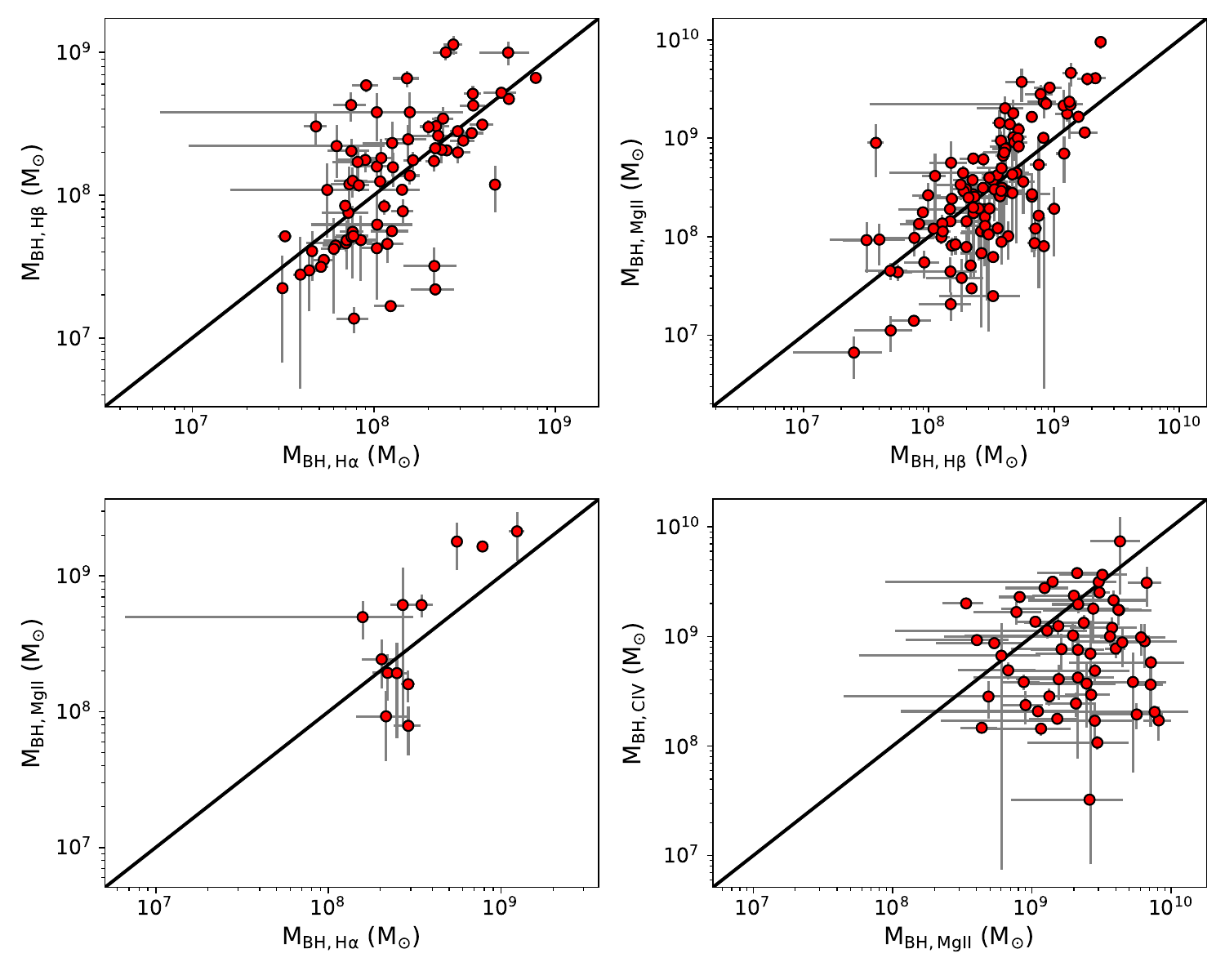}
  \caption{\label{comp_mbh} Intercomparison of black hole masses calculated from the various emission line single epoch formulas.We find good agreement between the Balmer-derived and \ion{Mg}{2}-derived black hole masses, albeit with a spread of a factor of 2-3, but a significant sysematic offset when comparing the \ion{Mg}{2}-derived black hole masses with those calculated from the \ion{C}{4} line. See Table \ref{bhm_residual} for details.}
\end{figure*}

We compared the agreement of the black hole mass measurements for the AGN in which we have these results from more than one emission line. As shown in Figure \ref{comp_mbh}, there is wide scatter when doing these pair-wise comparisons. We quantified the differences by calculating $\Delta$ = Log(M$_{\rm BH,1}$) - Log(M$_{\rm BH,2}$) and finding the average difference and dispersion in this residual, which we report in Table \ref{bhm_residual}. In all cases, the dispersion ranges from a factor of 2 - 4, consistent with the systematic uncertainties in black hole mass measurements reported in the literature \citep[i.e., 0.45 dex for H$\beta$ and \ion{Mg}{2}, 0.58 dex for \ion{C}{4};][]{shen2024}. On average, we find that the Balmer-derived and \ion{Mg}{2}-derived black hole masses agree with each other (with average offsets ranging from $\sim$2 - 20\%). However, the \ion{C}{4}-derived black hole masses are an average factor of 3 lower than those found from the \ion{Mg}{2} line. Due to this systematic difference between the \ion{Mg}{2} and \ion{C}{4} black hole masses, we recommend that statistical analyses using these black hole mass measurements focus on black hole masses measured using the {\it same} emission line single epoch formula. We note that a comparable study was done on moderate X-ray luminosity ($10^{43}$ erg s$^{1} < L_{\rm X} < 10^{45}$ eg s$^{-1}$) AGN from the COSMOS survey, which found comparable results: the black hole masses derived from the Balmer lines and the \ion{Mg}{2} line were consistent within factors of 0.01 - 0.06 dex with dispersions of 0.27 - 0.42 dex \citep{schulze2018}, though higher discrepancies were found when comparing black hole masses from the \ion{C}{4} line with those calculated from the H$\alpha$ and H$\beta$ lines (dispersions of 0.43 - 0.58 dex).

\begin{deluxetable}{lll}
\tablecaption{\label{bhm_residual}Offsets in Black Hole Mass Measurements}
\tablehead{\colhead{$M_{\rm BH}$ Residual} & \colhead{Mean Residual} & \colhead{Dispersion}  \\
  \colhead{Log($M_{\rm BH,1}$) - Log($M_{\rm BH,2}$)} & \colhead{(dex)} & \colhead{(dex)} 
}
\startdata
H$\beta$ - H$\alpha$ & 0.01 & 0.36 \\
\ion{Mg}{2} - H$\beta$ &  -0.01 & 0.43 \\
\ion{Mg}{2} - H$\alpha$ & 0.07 &  0.33 \\
\ion{C}{4} - \ion{Mg}{2} & -0.46 &  0.58 \\
\enddata
\end{deluxetable}

\begin{figure*}
 \includegraphics[scale=0.5]{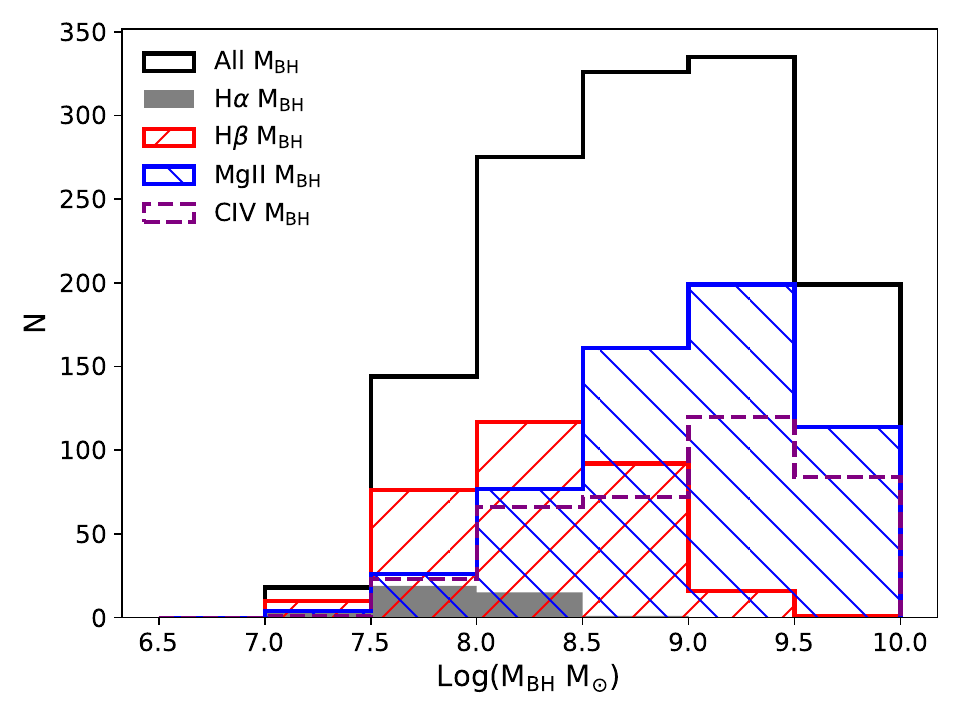}~
 \includegraphics[scale=0.5]{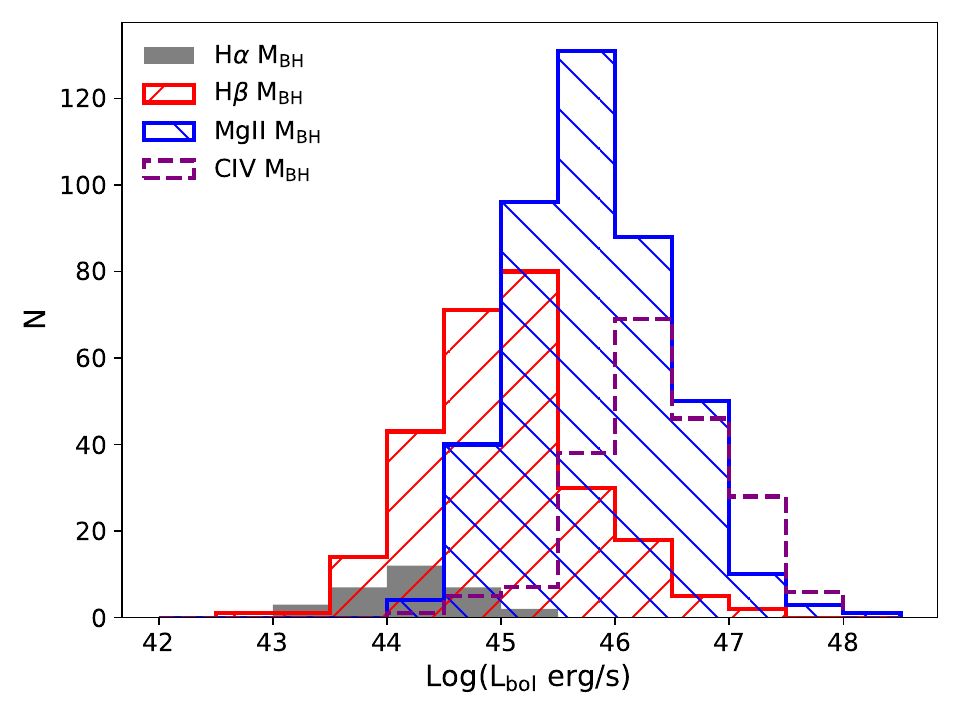}
 \caption{\label{comp_mbh_ref} ({\it Left}): Distribution of black hole masses for Type 1 AGN where the individual emission line single epoch calculations that are chosen as the reference $M_{\rm BH}$ are plotted separately. The higher mass black holes are found from the \ion{Mg}{2}- and \ion{C}{4}-derived black hole mass measurements. ({\it Right}): Distribution of the AGN bolometric luminosity for the various black hole mass formulations, showing that the AGN with \ion{Mg}{2}- and \ion{C}{4}-derived black hole masses are associated with more luminous AGN, which are also at higher redshift (see Table \ref{bhm_summary}).}
\end{figure*}

In Figure \ref{comp_mbh_ref} (left), we plot the distribution of the reference black hole masses (M$_{\rm BH}$) for the Stripe 82X Type 1 AGN, and highlight the subsets of measurements that are derived via the various emission line single epoch formulas used in this analysis. The AGN with more massive black holes are those calculated from the \ion{Mg}{2} and \ion{C}{4} emission lines. As we show in the right hand panel of Figure \ref{comp_mbh_ref}, the AGN with more massive black holes are more luminous (see Section \ref{ledd_sec} for $L_{\rm bol}$ calculation) and at higher redshift, consistent with the previously reported trends of AGN cosmic downsizing where more massive black holes power higher luminosity AGN at earlier times in the Universe ($z \gtrsim 1$) than in the present day \citep[$z \sim 0$; e.g.,][]{barger2005,ueda2014}.

\section{Eddington Luminosity and Ratios}\label{ledd_sec}
From the black hole mass, we calculate the Eddington Luminosity ($L_{\rm Edd}$), which is the luminosity at which radiation from the accretion disk balances gravity, with:
\begin{equation}
  L_{\rm Edd} = 1.26 \times 10^{38}\ \frac{M_{\rm BH}}{{\rm M_{\rm \odot}}}.
  \end{equation}
The value of $L_{\rm Edd}$ in the published catalog is based on the reference black hole mass ($M_{\rm BH}$).

We can calculate the Eddington parameter, which serves as a proxy of the accretion rate, by dividing the bolometric AGN luminosity ($L_{\rm bol}$) by the Eddington luminosity:
\begin{equation}
  \lambda_{\rm Edd} = L_{\rm bol}/L_{\rm Edd}
\end{equation}

We use the intrinsic X-ray luminosity calculated via spectral fitting in \citet{peca2023} to calculate $L_{\rm bol}$ using a bolometric luminosity dependent bolometric correction derived on a set of AGN that span almost 7 orders of magnitude in luminosity space \citep{duras2020}.

From \citet{duras2020}, the bolometric luminosity for Type 1 AGN is given as:
\begin{equation}\label{lbol_ty1}
\frac{L_{\rm bol}}{L_{\rm 2-10 keV}} = 12.76 \left[ 1 + \left(\frac{{\rm Log(L_{\rm bol}/L_{\rm \odot})}}{12.15}\right)^{18.87}\right].
\end{equation}

Out of our sample of 1297 Stripe 82X Type 1 AGN with measured black hole masses, 924 have measured intrinsic X-ray luminosity values from \citet{peca2023}, and thus $L_{\rm bol}$ values. We solved the above equation numerically to calculate $L_{\rm bol}$ for each AGN and use these values in the right hand panel of Figure \ref{comp_mbh_ref}.

To explore Eddington ratio trends, we are cognizant of the dispersion between Type 1 AGN black hole masses when using different single epoch formulas. We thus calculate the Eddington ratios separately for all AGN whose reference black hole mass is from the H$\beta$ formula ($\lambda_{\rm Edd,H\beta}$; 265 sources) and those derived from the \ion{Mg}{2} emission line ($\lambda_{\rm Edd,MgII}$; 506 sources). We choose both of these emission line single epoch measurements since as shown in Figure \ref{comp_mbh_ref}, these proxies span the full range of AGN bolometric luminosity and black hole mass values in this sample. 

\begin{figure*}
  \includegraphics[scale=0.5]{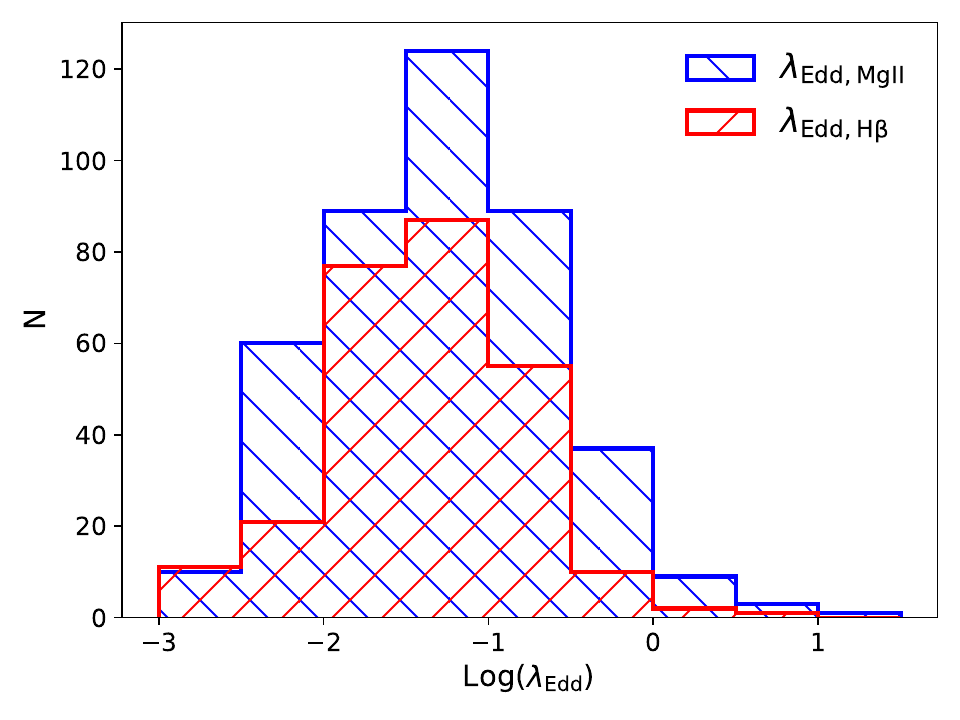}~
  \includegraphics[scale=0.5]{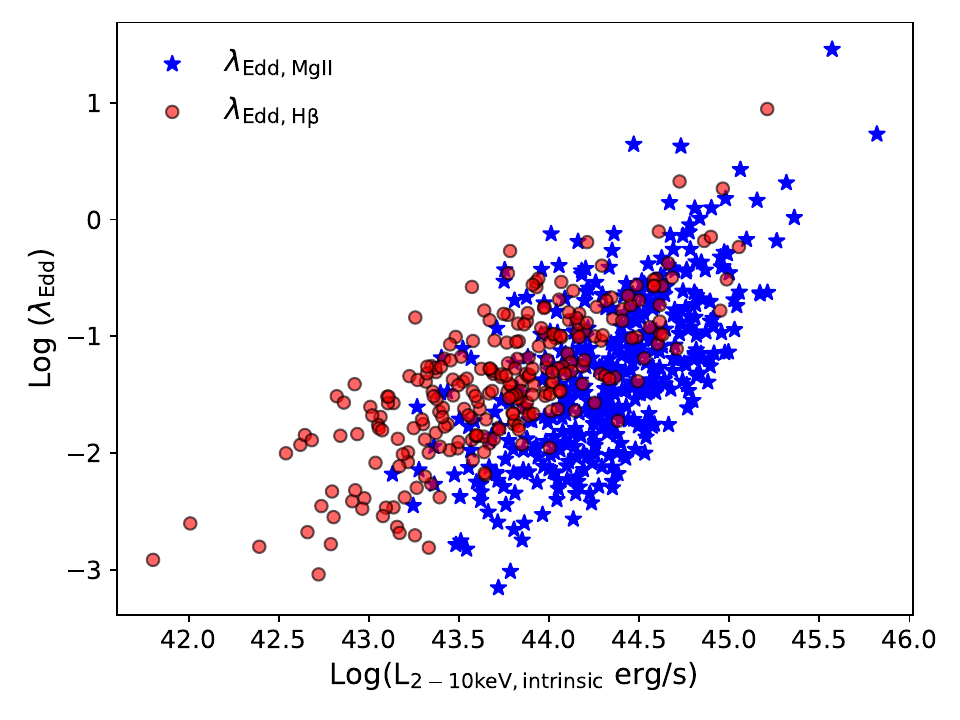}
  \includegraphics[scale=0.5]{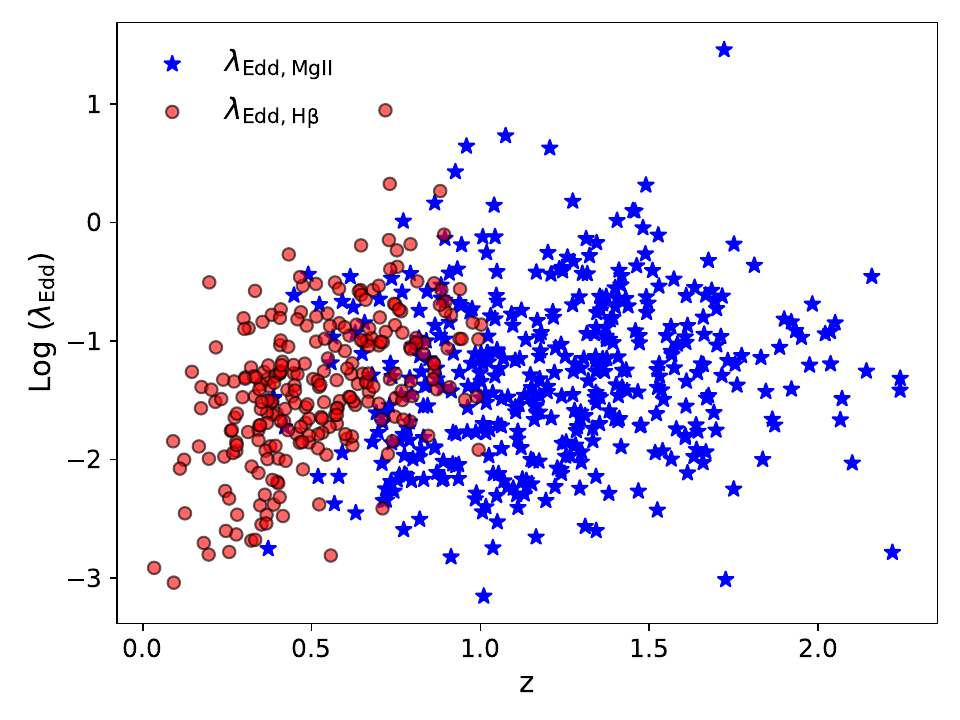}
  \caption{\label{edd_ratio} ({\it Left}): Distribution of the Eddington ratio ($\lambda_{\rm Edd}$) for Stripe 82X AGN that have intrinsic X-ray luminosities calculated via spectral fitting \citep{peca2023}. Due to the wide dispersion in black hole mass measurements from the broad emission line single epoch formulas, we consider separately Type 1 AGN with reference black hole mass measurements derived from the \ion{Mg}{2} line ($\lambda_{\rm Edd, MgII}$) and those derived from the H$\beta$ line ($\lambda_{\rm Edd, H\beta}$). Most Stripe 82X Type 1 AGN are accreting at rates below 30\% Eddington.
    ({\it Right}): Eddington ratio plotted as a function of the intrinsic X-ray luminosity (L$_{\rm 2-10keV,intrinsic}$) demonstrating that X-ray luminous AGN ($>10^{45}$ erg s$^{-1}$) are associated with higher accretion rates.
  ({\it Bottom}): Eddington ratio as a function of redshift, showing no strong trends with cosmic time, though the sample presented here is incomplete due to multiple selection effects.}
\end{figure*}

As shown in Figure \ref{edd_ratio}, there is a range of Eddington ratios, though most of the Stripe 82X Type 1 AGN are accreting at rates below 10-30\% Eddington. The right hand panel of Figure \ref{edd_ratio} demonstrates that these more rapidly accreting black holes ($\lambda_{\rm Edd} >$ 0.3) are in more X-ray luminous AGN, consistent with the framework that the bulk of black hole mass accretion occurs in a phase when the AGN is luminous \citep[e.g.,][]{kollmeier2006,hopkins2009,lusso2012}. We see no strong trends of Eddington ratio with AGN redshift (Figure \ref{edd_ratio}, bottom panel), though the sample presented here is incomplete due to multiple selection effects (i.e., X-ray flux limit of Stripe 82X, optical flux limit to identify counterparts, availability of SDSS spectra of adequate quality to measure $M_{\rm BH}$, and detection of an adequate number of X-ray photons for X-ray spectral fitting to determine the intrinsic X-ray luminosity from which the bolometric luminosity could then be estimated). We do see, though, that almost all super-Eddington AGN ($\lambda_{\rm Edd} > 1$) are at higher redshift ($z > 1$).

\section{Conclusions}
In this third data release of the 31.3 deg$^2$ Stripe 82X survey, we provide one comprehensive catalog that combines the X-ray, multiwavelength, and spectroscopic information published in previous releases \citep{lamassa2013a,lamassa2013b,lamassa2016a,lamassa2019,ananna2017} with 343 new spectroscopic redshifts. We also calculated black hole masses for 1297 Type 1 AGN.

Stripe 82X is 56\% spectroscopically complete: 3457 out of 6181 X-ray sources have secure spectroscopic redshifts. 93\% of the X-ray sources with redshifts are AGN, with $L_{\rm X} > 10^{42}$ erg s$^{-1}$. When considering the 20.2 deg$^2$ portion of the survey with homogenous X-ray coverage from {\it XMM-Newton} AO10 and AO13 (PI: Urry) that is brighter than $r = 22$, this completeness rises to 90\%. Focusing on this nearly complete portion of Stripe 82X, we find an obscured AGN fraction of 23\%, most of which are Type 2 AGN (i.e., they lack broad lines in their optical spectra). Unlike other surveys, we find only 18\% overlap between optically obscured and X-ray obscured ($N_{\rm H} > 10^{22}$ cm$^{-2}$) AGN.

Using single epoch spectroscopy virial mass formulas \citep{greene2005,shen2024}, we calculated black hole masses for 1297 Type 1 AGN. We find that the black hole masses calculated from the Balmer and \ion{Mg}{2} emission line single epoch formulas agree on average to within $\sim$2-20\%, while there is a systematic offset of a factor of $\sim$3 between the \ion{Mg}{2} and \ion{C}{4} $M_{\rm BH}$ values. The spread in the average values between the various formulas range from 0.33 - 0.58 dex. We encourage future statistical studies using Stripe 82X $M_{\rm BH}$ values to use black hole masses derived with the same formula. We found that most Stripe 82X AGN are accreting at sub-Eddington levels, and, consistent with previous studies, found that the most actively accreting black holes ($\lambda_{\rm Edd} > 1$) are hosted in the most X-ray luminous AGN ($L_{\rm 2-10 keV, intrinsic} > 10^{45}$ erg s$^{-1}$) at $z > 1$.

An upcoming Stripe 82X data release, Stripe 82-XL, will include additional archival X-ray observations, expanding the survey area to 55 deg$^2$ and totaling $\sim$23000 objects (Peca et al., accepted). We are currently in the process of releasing a morphological catalog for AGN host galaxies in Stripe 82X using the Galaxy Morphology Posterior Estimation Network \citep[GaMPEN;][]{ghosh2022} and PSFGAN \citep{tian2023}. A series of papers from the Accretion History of AGN collaboration (AHA, Co-PIs: C. M. Urry, D. Sanders) used this Stripe 82X dataset to explore the cosmic evolution of obscured, high X-ray luminosity AGN \citep{peca2023};  to model multi-wavelength spectral energy distributions (SEDs) of Stripe 82X AGN to determine SED evolution with X-ray luminosity \citep{auge2023}; and to study the star formation properties of Stripe 82X AGN host galaxies, leveraging the pre-existing {\it Herschel} data, to identify a new population of AGN dubbed ``cold quasars'' \citep{kirkpatrick2020} and to measure star formation rates and host galaxy stellar mass \citep{coleman2022}.

Stripe 82X, the forthcoming $\sim$55 deg$^2$ Stripe 82-XL (Peca et al., accepted) and other dedicted wide-area X-ray surveys like the 50 deg$^2$ {\it XMM}-XXL \citep{pierre2016,liu2016,menzel2016}, 13.1 deg$^2$ {\it XMM}- Spitzer Extragalactic Representative Volume Survey \citep[XMM-SERVS;][]{chen2018, ni2021}, and 9.3 deg$^2$ {\it Chandra} Bo{\"o}tes survey \citep{masini2020}, reveal the tip of the iceberg of luminous obscured black hole growth that can only be revealed by wide-area X-ray surveys. The eROSITA mission surveys the full sky in X-rays, with greatest sensivity in the soft 0.3-3.5 keV band \citep{predehl2021}, and can leverage lessons learned from dedicated surveys like Stripe 82X to more fully probe high redshift obscured AGN at high X-ray luminosity, which is a population we are only beginning to understand. Already, the eROSITA Final Equatorial Depth Survey (eFEDS), completed during the eROSITA performance verification phase, covered $\sim$140 deg$^2$ and detected over 27,000 significant X-ray sources \citep{brunner2022}, where 91\% were associated with a secure multiwavelength counterpart \citep{salvato2022} and 80\% were identied as AGN ($\sim$22,000 sources) based on their extragalactic spectroscopic or photometric redshifts \citep{liu2022}. While the obscured AGN fraction is modest (8\%, assuming a column density threshold of $N_{\rm H} = 10^{21.5}$ cm$^{-2}$ to differentiate between obscured and unobscured), eFEDS and the subsequent eROSITA All-Sky Survey \citep[eRASS1;][]{merloni2024}, culminating in a 4-year survey catalog \citep{predehl2021}, will provide valuable insight into obscured black hole growth at the highest X-ray luminosities.

\begin{acknowledgments}
  SML thanks D. Stern for advice when setting up the initial Palomar DoubleSpec and Keck LRIS observing runs, sharing IRAF routines to reduce the spectra, and for observing a handful of Stripe 82X sources as fillers in other observing programs from which spectroscopic redshifts and classifications were obtained (DBSP\_2014\_Dec, DBSP\_2017\_Jul, DBSP\_2017\_Sep, Keck\_LRIS\_Sep2017).
  TTA acknowledges support from ADAP grant 80NSSC23K0557.
 Some of the data presented herein were obtained at Keck Observatory, which is a private 501(c)3 non-profit organization operated as a scientific partnership among the California Institute of Technology, the University of California, and the National Aeronautics and Space Administration. The Observatory was made possible by the generous financial support of the W. M. Keck Foundation.
  The authors wish to recognize and acknowledge the very significant cultural role and reverence that the summit of Maunakea has always had within the Native Hawaiian community. We are most fortunate to have the opportunity to conduct observations from this mountain.
  This manuscript uses spectroscopic redshifts from SDSS-IV, SDSS-III, and SDSS Legacy. Funding for the Sloan Digital Sky Survey IV has been provided by the Alfred P. Sloan Foundation, the U.S. Department of Energy Office of Science, and the Participating Institutions. SDSS acknowledges support and resources from the Center for High-Performance Computing at the University of Utah. The SDSS web site is www.sdss4.org.
SDSS is managed by the Astrophysical Research Consortium for the Participating Institutions of the SDSS Collaboration including the Brazilian Participation Group, the Carnegie Institution for Science, Carnegie Mellon University, Center for Astrophysics | Harvard \& Smithsonian (CfA), the Chilean Participation Group, the French Participation Group, Instituto de Astrofísica de Canarias, The Johns Hopkins University, Kavli Institute for the Physics and Mathematics of the Universe (IPMU) / University of Tokyo, the Korean Participation Group, Lawrence Berkeley National Laboratory, Leibniz Institut f\"ur Astrophysik Potsdam (AIP), Max-Planck-Institut f\"ur Astronomie (MPIA Heidelberg), Max-Planck-Institut f\"ur Astrophysik (MPA Garching), Max-Planck-Institut f\"ur Extraterrestrische Physik (MPE), National Astronomical Observatories of China, New Mexico State University, New York University, University of Notre Dame, Observat\'orio Nacional / MCTI, The Ohio State University, Pennsylvania State University, Shanghai Astronomical Observatory, United Kingdom Participation Group, Universidad Nacional Autónoma de M\'exico, University of Arizona, University of Colorado Boulder, University of Oxford, University of Portsmouth, University of Utah, University of Virginia, University of Washington, University of Wisconsin, Vanderbilt University, and Yale University.
\end{acknowledgments}

%

\vspace{5mm}
\facilities{XMM, CXO, Keck:I (LRIS), Keck:II (DEIMOS), Keck:II (NIRES), Keck:II (NIRSPEC), Hale (DBSP), Hale (TripleSpec), WIYN (HYDRA), Gemini:Gillett (GNIRS), GALEX, Sloan, ESO:VISTA, UKIRT, Spitzer, WISE, NEOWISE, VLA, Herschel}


\software{astropy \citep{astropy2013,astropy2018,astropy2022},
  PypeIt \citep{prochaska2020a, prochaska2020b},
SPEXTOOL \citep{cushing2004} }

\appendix

\section{Stripe 82X Catalog Description}

This Stripe 82X catalog release combines elements of Stripe 82X Data Release 1 \citep[X-ray source properties, far-infrared and radio counterparts to X-ray sources;][]{lamassa2016a}, Stripe 82X Data Release 2 \citep[multiwavelength counterpart coordinates and photometry, and photometric redshifts;][]{ananna2017}, spectroscopic redshifts from Stripe 82X SDSS-IV eBOSS \citep{lamassa2019}, new spectroscopic redshifts and classifications, and the black hole masses calculated in this paper. We utilize data from the {\it Chandra} Source Catalog \citep{evans2010} for the archival {\it Chandra} sources in Stripe 82X \citep{lamassa2013a}, where we required that a source be detected at $\geq$4.5$\sigma$ in at least one X-ray energy band (soft, hard, or full) for inclusion in the catalog. We performed source detection on all the {\it XMM}-Newton observations in Stripe 82X, both archival sources \citep{lamassa2013b} and the observations awarded to us in AO10 \citep{lamassa2013b} and AO13 \citep{lamassa2016a}. We quantified the significance of an X-ray source detection using a minimum likelihood threshold: {\it det\_ml} = -ln$P_{\rm random}$, where $P_{\rm random}$ is the Poissonian probability that an X-ray source detection is due to chance. To include an {\it XMM} source in the catalog, it had to be detected at a {\it det\_ml} $\geq$ 15 (i.e., $>5\sigma$) level in at least one X-ray band.

When independently matching multiwavelength catalogs to the X-ray source list using a maximum likelihood estimator approach \citep{sutherland1992}, \citet{ananna2017} defined a threshold likelihood ratio (LR) for each wavelength band above which they consider a match reliable. They also calculated a reliability value for each association following the procedure in \citet{marchesi2016a}. They found that different multiwavelength bands chose different counterparts for 12\% of X-ray sources. In their analysis, they identified the most likely counterpart to each X-ray source and included the alternate counterparts as a reference, defining a quality flag (``QF'') to describe the conflicts. In this catalog, we chose the most likely counterpart and report the ``QF'' value for reference so users can assess the reliability of the association. All magnitudes are reported in the AB system. \citet{ananna2017} used this multiwavelength catalog to generate spectral energy distributions (SEDs) that they fitted with \textsc{LePhare} \citep{lephare} using a range of input SED templates to calculate photometric redshifts.

We describe the contents of the Stripe 82X catalog in Table \ref{cat_columns}. In Table \ref{specz_cat}, we list the number of Stripe 82X sources whose spectroscopic redshifts were collected from surveys and archival catalogs.

\begin{figure}
  \center
  \includegraphics[scale=0.6]{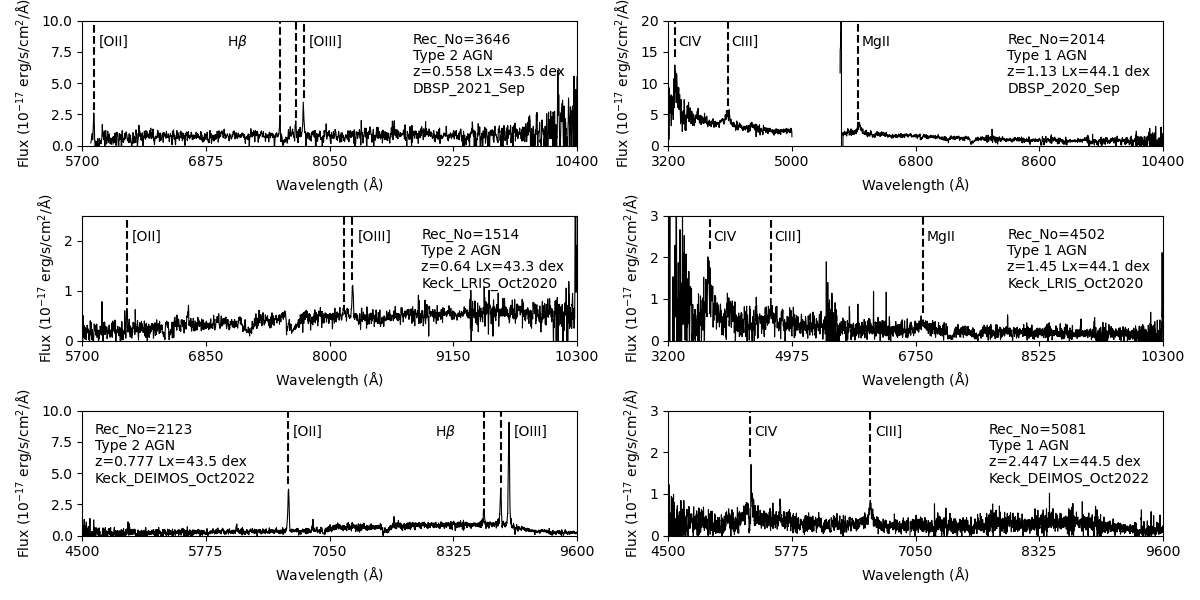}
  \caption{\label{opt-spectra} Example of spectra from our follow-up campaign with (top) Palomar DoubleSpec, (middle) Keck LRIS, and (bottom) Keck DEIMOS, illustrating both (left) Type 2 AGN and (right) Type 1 AGN. Identifying information is given in the legend of each plot, including estimated 2-10 keV luminosity and observing run in which spectrum was taken. Emission lines used to determine spectroscopic redshift are marked. Spectra from the red (blue) arm are Gaussian smoothed with a $\sigma$=1.0 ($\sigma$=1.5) kernel. The spectra are not telluric corrected nor do they have an absolute flux calibration applied.}
\end{figure}

In Table \ref{specz_our_campaign}, we list the number of Stripe 82X sources with redshifts and classifications from our follow-up spectroscopic campaign. Details about the observing set-ups for our ground-based campaign are given in Table \ref{obs_setup}. Spectra obtained prior to 2018 were reduced with IRAF or SPEXTOOL \citep[for Palomar TripleSpec spectra;][]{cushing2004,lamassa2017} while infrared (optical) spectra obtained after 2018 (2020) were reduced with PypeIt \citep{prochaska2020a,prochaska2020b}. The spectroscopic redshifts published here for the first time are those obtained after 2015 September, with the exception of the infrared spectra from Gemini GNIRS (2015) and Palomar TripleSpec (2015) which were published in \citet{lamassa2017}. A handful of optical spectra from our follow-up are shown in Figure \ref{opt-spectra} for illustrative purposes.

\startlongtable
\begin{deluxetable}{p{0.25\linewidth}  p{0.6\linewidth}}
  \tablecaption{\label{cat_columns}Descripton of Stripe 82X Catalog Columns}
  \tablehead{\colhead{Column Name} & \colhead{Description}}
  \startdata
  MSID & {\it Chandra} Source Catalog \citep{evans2010} identifier; set to 0 for {\it XMM sources}. \\
  REC\_No & Stripe 82X identifier number for archival and AO10 and AO13 {\it XMM} sources; set to 0 for {\it Chandra} sources. \\
  ObsID & {\it XMM} or {\it Chandra} Observation ID number for observation in which source is detected \\
  Xray\_Src & Provenance of X-ray source. Possible values are ``Chandra'', ``XMM\_archive'', ``XMM\_AO10'', or ``XMM\_AO13''. \\
  Xray\_RA, Xray\_Dec & Right Ascension and Declination of X-ray source. \\
  Xray\_RADec\_Err & Astrometric error on the X-ray source coordinates in arcseconds. \\
  Soft\_Flux, Hard\_Flux, Full\_Flux & Observed flux (erg s$^{-1}$ cm$^{-2}$) in the soft (0.5-2 keV), hard (2-10 keV for {\it XMM}; 2-7 keV for {\it Chandra}), and full (0.5-10 keV for {\it XMM}; 0.5-7 keV for {\it Chandra}) bands.  A powerlaw model with $\Gamma = 2.0$ ($\Gamma = 1.7$) is assumed to convert count rates to flux for the soft (hard and full) bands. Users are advised to check {\it det\_ml} for significance of flux measurement. \\
  Soft\_Flux\_Error, Hard\_Flux\_Error, Full\_Flux\_Error & Error on the soft, hard, and full band fluxes (erg s$^{-1}$ cm$^{-2}$). Values are for {\it XMM} sources only. \\
  Soft\_Flux\_Error\_High, Hard\_Flux\_Error\_High, Full\_Flux\_Error\_High  & Higher bound on the soft, hard, and full band fluxes (erg s$^{-1}$ cm$^{2}$). If corresponding flux measurements are zero, these values represent the upper limit. Values are for {\it Chandra} sources only. \\
  Soft\_Flux\_Error\_Low, Hard\_Flux\_Error\_Low, Full\_Flux\_Error\_Low  & Lower bound on the soft, hard, and full band fluxes (erg s$^{-1}$ cm$^{-2}$). Values are for {\it Chandra} sources only. \\
  Soft\_Counts, Hard\_Counts, Full\_Counts & Net counts in the soft, hard, and full bands.\\
  Soft\_DetML, Hard\_DetML, Full\_DetML & Significance of detection in soft, hard, and full band, given by ${\it det\_ml}=-{\rm ln}P_{\rm random}$. For all significant {\it Chandra} detections, {\it det\_ml} is set to a fiducial value of 11.0439, representing a 4.5$\sigma$ detection which we required to include a source in the LogN-LogS analysis in \citet{lamassa2013a,lamassa2016a}. For the {\it XMM} sources, we required {\it det\_ml} $\geq$15 to include a source in the LogN-LogS analysis of \citet{lamassa2013b,lamassa2016a}. \\
  Soft\_Lum, Hard\_Lum, Full\_Lum & $k$-corrected luminosity in the soft, hard, and full bands, calculated for sources with a spectroscopic redshift and where {\it det\_ml} $>$10 in that band. Units are in erg s$^{-1}$. Luminosities are reported in log space. \\
  CP\_RA, CP\_Dec & Coordinates of multiwavelength counterpart to X-ray source. Coordinates may be from SDSS, then VHS, then {\it Spitzer}. See \citet{ananna2017} for details. \\
  QF & Quality flag of multiwavelength counterpart (see \citet{ananna2017} for details):\\
  & 0 No association found, or counterpart is from the \citet{lamassa2016a} catalog. \\
  & 1 Unambiguous association. \\
  & 1.5 Conflicting associations, but a secure match in one band helps to resolve ambiguity. \\
  & 2 Counterparts disagree among multiwavelength bands matched to X-ray source, but the counterpart in one band exceeds the LR threshold to accept the association as the true counterpart, while the LR value was below the threshold in the other band. \\
  & 3 Counterparts disagree among multiwavelength bands and the LR values are comparable. The association with the greater reliability value is chosen as the counterpart. \\
  & 4 A counterpart is found in only one multiwavelength band. \\
  mag\_FUV, magerr\_FUV, mag\_NUV, magerr\_NUV & {\it GALEX} far ultraviolet (FUV) and near ultraviolet (NUV) magnitudes and errors \citep[AB;][]{morrissey2007}. Extinction-corrected. \\
  u, u\_err, g, g\_err, r, r\_err, i, i\_err, z, z\_err & Optical SDSS magnitudes and errors (AB) from the coadded catalogs of \citet{fliri2016} (if available), otherwise from \citet{jiang2014}. Extintion-corrected when extinction information is available.\\
  JVHS, JVHS\_err, HVS, HVHS\_err, HVHS\_err, KVHS, KVHS\_err & Near-infrared magnitudes and errors (AB) from the VISTA VHS survey in the $J$, $H$, and $K$ bands, using a 2.8$^{\prime\prime}$ (5.6$^{\prime\prime}$) radius aperture for point (extended) sources \citep{mcmahon2013}. Extinction-corrected. \\
  JUK, JUK\_err, HUK, HUK\_err, HUK\_err, KUK, KUK\_err & Near-infrared magnitudes and errors (AB) from the UKIDSS Large Area Survey survey in the $J$, $H$, and $K$ bands, using a 2.8$^{\prime\prime}$ (5.6$^{\prime\prime}$) radius aperture for point (extended) sources \citep{lawrence2007}. Extinction-corrected. \\
  CH1\_SpIES, CH1\_SpIES\_err, CH2\_SpIES, CH2\_SpIES\_err & {\it Spitzer} IRAC channel 1 (3.6 $\mu$m) and channel 2 (4.5 $\mu$m) magnitudes and error (AB) from the SpIES survey \citep{timlin2016}. Not extinction-corrected. \\
  CH1\_SHELA, CH1\_SHELA\_err, CH2\_SHELA, CH2\_SHELA\_err & {\it Spitzer} IRAC channel 1 (3.6 $\mu$m) and channel 2 (4.5 $\mu$m) magnitudes and error (AB) from the SHELA survey \citep{papovich2016}. Not extinction-corrected. \\
 W1, W1\_err, W2, W2\_err, W3, W3\_err, W4, W4\_err & AllWISE \citep{allwise} instrumental profile-fit photometry (w$n$mpro, where $n$ refers to band number) and error for point sources or elliptical aperture magnitude and error (w$n$gmag) for extended sources (where WISE\_Ext is set to ``YES''). W1, W2, W3, and W4 refer to magnitudes at 3.4$\mu$m, 4.6$\mu$m, 12$\mu$m, and 22$\mu$m, respectively. Reported magnitudes in this catalog are in AB system. Not extinction-corrected.\\
 W1\_SNR, W2\_SNR, W3\_SNR, W4\_SNR & Signal-to-noise ratio for each {\it WISE} band. {\it WISE} magnitudes with S/N $<$ 2 should be considered an upper limit.\\
 W1\_Sat, W2\_Sat, W3\_Sat, W4\_Sat & Fraction of pixels affected by saturation in each {\it WISE} band.\\
 WISE\_Ext & Flag to indicate if {\it WISE} source is extended. Reported magnitudes refer to elliptical magnitudes if set to ``YES.''\\
 FIRST\_RA, FIRST\_Dec & Coordinates of radio counterpart from FIRST survey \citep{white1997}. \\
 FIRST\_Dist & Distance (in arcseconds) between X-ray source and FIRST counterparts. Due to the low space density of FIRST sources, counterparts were found using nearest neighbor matching in \citet{lamassa2016a}.\\
 FIRST\_Flux, FIRST\_Flux\_Err & Integrated flux density and error at 1.4 GHz (mJy) from FIRST survey \citep{white1997}. \\
 HERS\_RA, HERS\_Dec & Coordinates of far-infrared {\it Herschel} counterpart to X-ray source \citep{viero2014}.\\
 HERS\_Dist &  Distance (in arcseconds) between X-ray source and {\it Herschel} counterparts. Due to the low space density of {\it Herschel} sources, counterparts were found using nearest neighbor matching in \citet{lamassa2016a}.\\
 F250, F250\_Err, F350, F350\_Err, F500, F500\_Err & Flux density and $1\sigma$ error at 250$\mu$m, 350$\mu$m, and 500$\mu$m (mJy) from {\it Herschel} HeRS survey \citep{viero2014}.\\
 spec\_z & Spectroscopic redshift. \\
 spec\_class & Spectroscopic classification. \\
 z\_src & Source of spectroscopic redshift and classification. See Table \ref{specz_cat} for spectroscopic information from surveys and archival catalogs and Tables \ref{specz_our_campaign}  - \ref{obs_setup} for spectroscopic information from our ground-based follow-up campaign.\\
 photo\_z & Best-fit photometric redshift (see \citet{ananna2017} for details). \\
 PDZ\_best & Confidence on photometric redshift. \\
 Photo\_z\_sec & Secondary photometric redshift with a fit quality close to that of photo\_z.\\
 PDZ\_sec & Confidence on secondary photometric redshift. \\
 SED\_morphology & Morphology of source according to SED template fitting for calculating photo\_z with \textsc{LePhare} \citep{lephare}:
 \begin{enumerate}
 \item Star
 \item Elliptical galaxy
 \item Spiral galaxy
 \item Type 2 AGN
 \item Starburst galaxy
   \item Type 1 AGN
   \item QSO
   \end{enumerate} \\
 Morphology & Source morphology according to optical or NIR photometry: point-like, extended, or uncertain \citep[see][]{ananna2017}. \\  
 M\_BH, M\_BH\_err & Reference black hole mass and error in units of M$_{\rm \odot}$.\\
 M\_BH\_Src & Emission line single epoch formula used to derive M\_BH. We used the following ranked ordering as described in the main text:
 \begin{enumerate}
 \item H$\beta$ emission line formula from \citet{shen2024};
 \item MgII emission line formula from \citet{shen2024};
 \item H$\alpha$ emission line formula from \citet{greene2010};
 \item CIV emission line formula from \citet{shen2024};
 \end{enumerate}\\
L\_Edd, L\_Edd\_err & Eddington luminosity and error calculated from reference M\_BH. \\
 M\_BH\_HB, M\_BH\_HB\_err, M\_BH\_MgII, M\_BH\_MgII\_err,  M\_BH\_Ha, M\_BH\_Ha\_err, M\_BH\_CIV, M\_BH\_CIV\_err & Black hole masses and errors derived from the H$\beta$ \citep{shen2024}, MgII \citep{shen2024}, H$\alpha$ \citep{greene2010}, and CIV \citep{shen2024} formulas, respectively, in units of M$_{\rm \odot}$. Due to the wide dispersion when comparing black hole masses from the same AGN using different formulas, we recommend that statistical analyses use black hole masses calculated using the same formula.\\
 Xray\_Lum & Estimate of the $k$-corrected, non absorprtion corrected hard band (2-10 keV) X-ray luminosity from the X-ray count rate. This value represents the hard band X-ray luminosity if there is a significant detection in the hard band. Otherwise, the X-ray luminosity in the full or soft band is scaled to derive $L_{\rm X}$, as described in the main text. This value is only calculated for sources with spectroscopic redshifts that are extragalactic. Luminosity is in units of erg s$^{-1}$ and is reported in log space. For intrinsic X-ray luminosities derived via X-ray spectral fitting, see \citet{peca2023}. \\
 Xray\_AGN & If Xray\_Lum exceeds 10$^{42}$ erg s$^{-1}$, this flag is set to ``True.''
  \enddata
\end{deluxetable}

\begin{deluxetable}{p{0.25\linewidth} p{0.1\linewidth}  p{0.5\linewidth}}
\tablecaption{\label{specz_cat}Spectroscopic Redshifts from Surveys and Archival Catalogs}
\tablehead{\colhead{Catalog ID\tablenotemark{a}} & \colhead{Number} & \colhead{Reference}}
\startdata
2SLAQ  & 67 & 2dF-SDSS QSO survey \citep{croom2009} \\
6dF & 6 & 6dF Galaxy Survey \citep{jones2004,jones2009} \\
DEEP2 & 9 & DEEP2 survey \citep{newman2013} \\
eBOSS\_S82X & 616 & SDSS-IV eBOSS survey of Stripe 82X \citep{lamassa2019} \\
Jiang2006 & 4 & Spectroscopic survey of faint quasars in Stripe 82 \citep{jiang2006} \\
pre-BOSS & 6 & pre-SDSS BOSS survey \citep{ross2012} \\
PRIMUS & 12 & PRIMUS survey \citep{coil2011} \\
SDSS\_DR7Q & 46 & SDSS Quasar Catalog, Seventh Data Release \citep{dr7q}\\
SDSS\_DR8 & 4 & Eighth Data Release of SDSS \citep{sdssdr8} \\
SDSS\_DR9 & 4 & Ninth Data Release of SDSS \citep{sdssdr9} \\
SDSS\_DR12 & 43 & Eleventh and Twelfth Data Releases of SDSS \citep{sdssdr12}\\
SDSS\_DR12Q & 511 & SDSS Quasar Catalog, Twelfth Data Release \citep{dr12q}\\
SDSS\_DR13 & 975 & Thirteenth Data Release of SDSS \citep{sdssdr13} \\
SDSS\_DR14 & 33 & Fourteenth Data Release of SDSS \citep{sdssdr14} \\
SDSS\_DR14Q & 490 & SDSS Quasar Catalog, Fourteenth Data Release \citep{dr14q} \\
SDSS\_DR16 & 107 & Sixteenth Data Release of SDSS \citep{sdssdr16} \\
sdss\_correction & 2 & SDSS spectra where the reported redshift was incorrect and we report the correct redshift based on an independent calculation \\
SDSS\_zwarning\_verified\_by\_eye & 13 & SDSS spectra where the \textsc{z\_warning} flag was set but we verified via visual inspection that the redshift and classification are acceptable \\
VVDS & 24 & Vimos VLT Deep Survey \citep{lefevre2004, lefevre2005, lefevre2013, garilli2008} \\
WiggleZ & 7 &WiggleZ survey \citep{drinkwater2010}
\enddata
\tablenotetext{a}{Identifier used in ``z\_src'' column of Stripe 82X catalog.}
\end{deluxetable}

\begin{deluxetable}{lll}
\tablecaption{\label{specz_our_campaign}Spectroscopic Redshifts from Dedicated Follow-up Campaigns}
\tablehead{\colhead{Catalog ID\tablenotemark{a}} & \colhead{Number} & \colhead{UT Observing Date}}
\startdata
HYDRA\_2012\_Dec & 14 & 2012 Dec 12 - Dec 13; 2012 Dec 17\\
HYDRA\_2013\_Aug\_Sep & 22 & 2013 Aug 31 - Sep 2\\
HYDRA\_2014\_Jan & 44 & 2014 Jan 5 - Jan 7 \\
HYDRA\_2014\_Jun & 12 & 2014 Jun 21 - Jun 22 \\
HYDRA\_2014\_Jul & 29 & 2014 Jul 24 - Jul 28 \\
DBSP\_2014\_Jul &  4 & 2024 Jul 23 - Jul 24 \\
Keck\_NIRSPEC\_2014\tablenotemark{b} & 1 & 2014 Sep 7 \\
DBSP\_2014\_Dec &  1 & 2014 Dec 23 \\
HYDRA\_2015\_Jan & 1 & 2015 Jan 15 - Jan 16 \\
DBSP\_2015\_Sep & 22 & 2015 Sep 8 - Sep 11 \\
DBSP\_2015\_Oct &  7 & 2015 Oct 13 - Oct 17 \\
Gemini\_GNIRS\_2015\tablenotemark{b} & 3 & 2015 Dec 10; 2016 Jan 2; 2016 Jan 5; 2016 Jan 7; 2016 Jan 8\\
Palomar\_TSpec\_2015\tablenotemark{b} & 4 & 2015 Oct 27 \\
DBSP\_2016\_Aug & 25 & 2016 Aug 28 - Aug 31 \\
DBSP\_2016\_Sep & 42 & 2017 Sep 24 - Sep 28 \\
DBSP\_2017\_Jul & 2 & 2017 Jul 27 - Jul 28 \\
DBSP\_2017\_Aug &  18 & 2017 Aug 27 - Aug 28 \\
DBSP\_2017\_Sep & 1 & 2017 Sep 1 \\
Keck\_LRIS\_Sep2017 & 1 & 2017 Sep 16 \\
Keck\_LRIS\_Oct2017 & 16 & 2017 Oct 17 - Oct 18\\
DBSP\_2017\_Oct & 17 & 2017 Oct 27 - 28\\
Keck\_NIRES\_Sep2018 & 6 & 2018 Sep 17; 2018 Sep 19 \\
Gemini\_GNIRS\_2019 & 1 & 2019 Sep 11; 2019 Oct 25 \\
Keck\_NIRES\_Oct2019 & 8 & 2019 Oct 15\\
Palomar\_TSpec\_Oct2019 & 3 & 2019 Oct 16 - Oct 21 \\
DBSP\_2020\_Sep & 18 & 2020 Sep 23 - Sep 24 \\
Keck\_NIRES\_Oct2020 & 5 & 2020 Oct 7 \\
Keck\_LRIS\_Oct2020 & 17 & 2020 Oct 11\\
DBSP\_2021\_Jul & 23 & 2021 Jul 7 - Jul 8; 2021 Jul 16 - Jul 18\\
DBSP\_2021\_Sep & 30 & 2021 Sep 28 - Oct 1 \\
KECK\_DEIMOS\_Oct2022 & 15 & 2022 Oct 3 - Oct 4 \\
Keck\_NIRES\_Oct2022 & 1 & 2022 Oct 5 \\
DBSP\_2022\_Oct & 65 & 2022 Oct 4 - Oct 5;  2022 Oct 17 - Oct 20
\enddata
\tablenotetext{a}{Identifier used in ``z\_src'' column of Stripe 82X catalog. See Table \ref{obs_setup} for details.}
\tablenotetext{b}{See \citet{lamassa2017}.}
\end{deluxetable}

\begin{deluxetable}{llllll}
  \tablecaption{\label{obs_setup}Instrument Set-up for Follow-up Observing Campaigns}
  \tablehead{\colhead{``z\_src'' prefix} &\colhead{Observatory} & \colhead{Instrument} & \colhead{Wavelength Coverage} & \colhead{Notes} & \colhead{References}}
  \startdata
  \multicolumn{6}{c}{Optical Instruments}\\
  \hline
  HYDRA & WIYN & HYDRA & 3000 - 8000 $\rm{\AA}$ & Multi-object spectrograph & \citet{barden1995} \\
  DBSP & Palomar & DoubleSpec & 3200 - 5500 $\rm{\AA}$  (blue) & Single slit; & \citet{oke1982}\\
  & & & 4700 - 10500 $\rm{\AA}$  (red) & separate blue and red arms & \citet{rahmer2012} \\
 KECK\_LRIS & Keck & Low Resolution & 3150 - 5400 $\rm{\AA}$  (blue) & Single slit; & \citet{oke1995} \\
 & & Imaging Spectrometer & 5650 - 10300 $\rm{\AA}$ (red) & separate blue \& red arms & \citet{rockosi2010} \\
 KECK\_DEIMOS & Keck & DEep Imaging Multi- & 4500 - 6990 $\rm{\AA}$ (blue) & Used in single slit mode; & \citet{faber2003} \\
 & & Object Spectrograph & 7000 - 9630 $\rm{\AA}$ (red) & separate blue \& red arms & \\
 \hline
 \multicolumn{6}{c}{Infrared Instruments}\\
  \hline
 Keck\_NIRSPEC & Keck & NIRSPEC & 1.95 - 2.3 $\mu$m & Echelle & \citet{mclean1998} \\
 Palomar\_TSpec & Palomar & Triple Spectrograph & 1.0 - 2.4 $\mu$m & Echelle & \citet{herter2008} \\
 Gemini\_GNIRS & Gemini & Gemini Near- & 0.8 - 2.5 $\mu$m & Echelle & \citet{elias1998} \\
 & & Infrared Spectrograph \\
 Keck\_NIRES & Keck & Near-Infrared Echellete &  0.94 - 2.45 $\mu$m & Echelle & \citet{wilson2004} \\
  & & Spectrometer & & & \citet{kassis2018} 
   \enddata
   \tablenotetext{}{Observations with Palomar used the 200-inch Hale Telescope.}
 \end{deluxetable}

\begin{thebibliography}

\bibitem[Abazajian et al.(2009)]{sdssdr7} Abazajian, K.~N., Adelman-McCarthy, J.~K., Ag{\"u}eros, M.~A., et al.\ 2009, \apjs, 182, 543. doi:10.1088/0067-0049/182/2/543

\bibitem[Abolfathi et al.(2018)]{sdssdr14} Abolfathi, B., Aguado, D.~S., Aguilar, G., et al.\ 2018, \apjs, 235, 42. doi:10.3847/1538-4365/aa9e8a

\bibitem[Ahn et al.(2012)]{sdssdr9} Ahn, C.~P., Alexandroff, R., Allende Prieto, C., et al.\ 2012, \apjs, 203, 21. doi:10.1088/0067-0049/203/2/21

\bibitem[Ahumada et al.(2020)]{sdssdr16} Ahumada, R., Allende Prieto, C., Almeida, A., et al.\ 2020, \apjs, 249, 3. doi:10.3847/1538-4365/ab929e

\bibitem[Aihara et al.(2011)]{sdssdr8} Aihara, H., Allende Prieto, C., An, D., et al.\ 2011, \apjs, 193, 29. doi:10.1088/0067-0049/193/2/29

\bibitem[Aird et al.(2015)]{aird2015} Aird, J., Coil, A.~L., Georgakakis, A., et al.\ 2015, \mnras, 451, 1892. doi:10.1093/mnras/stv1062

\bibitem[Alam et al.(2015)]{sdssdr12} Alam, S., Albareti, F.~D., Allende Prieto, C., et al.\ 2015, \apjs, 219, 12. doi:10.1088/0067-0049/219/1/12

\bibitem[Albareti et al.(2017)]{sdssdr13} Albareti, F.~D., Allende Prieto, C., Almeida, A., et al.\ 2017, \apjs, 233, 25. doi:10.3847/1538-4365/aa8992

\bibitem[Antonucci(1993)]{antonucci1993} Antonucci, R.\ 1993, \araa, 31, 473. doi:10.1146/annurev.aa.31.090193.002353
  
\bibitem[Ananna et al.(2017)]{ananna2017} Ananna, T.~T., Salvato, M., LaMassa, S., et al.\ 2017, \apj, 850, 66. doi:10.3847/1538-4357/aa937d

\bibitem[Ananna et al.(2019)]{ananna2019} Ananna, T.~T., Treister, E., Urry, C.~M., et al.\ 2019, \apj, 871, 240. doi:10.3847/1538-4357/aafb77
  
\bibitem[Annis et al.(2014)]{annis2014} Annis, J., Soares-Santos, M., Strauss, M.~A., et al.\ 2014, \apj, 794, 120. doi:10.1088/0004-637X/794/2/120

\bibitem[Arnouts \& Ilbert(2011)]{lephare} Arnouts, S. \& Ilbert, O.\ 2011, Astrophysics Source Code Library. ascl:1108.009

\bibitem[Arrabal Haro et al.(2023)]{arrabalharo2023} Arrabal Haro, P., Dickinson, M., Finkelstein, S.~L., et al.\ 2023, \apjl, 951, L22. doi:10.3847/2041-8213/acdd54

\bibitem[Astropy Collaboration et al.(2013)]{astropy2013} Astropy Collaboration, Robitaille, T.~P., Tollerud, E.~J., et al.\ 2013, \aap, 558, A33. doi:10.1051/0004-6361/201322068

\bibitem[Astropy Collaboration et al.(2018)]{astropy2018} Astropy Collaboration, Price-Whelan, A.~M., Sip{\H{o}}cz, B.~M., et al.\ 2018, \aj, 156, 123. doi:10.3847/1538-3881/aabc4f

 \bibitem[Astropy Collaboration et al.(2022)]{astropy2022} Astropy Collaboration, Price-Whelan, A.~M., Lim, P.~L., et al.\ 2022, \apj, 935, 167. doi:10.3847/1538-4357/ac7c74

 \bibitem[Auge et al.(2023)]{auge2023} Auge, C., Sanders, D., Treister, E., et al.\ 2023, \apj, 957, 19. doi:10.3847/1538-4357/acf21a

 \bibitem[Barden \& Armandroff(1995)]{barden1995} Barden, S.~C. \& Armandroff, T.\ 1995, \procspie, 2476, 56. doi:10.1117/12.211839

\bibitem[Baron \& M{\'e}nard(2019)]{baron} Baron, D. \& M{\'e}nard, B.\ 2019, \mnras, 487, 3404. doi:10.1093/mnras/stz1546

\bibitem[Barger et al.(2005)]{barger2005} Barger, A.~J., Cowie, L.~L., Mushotzky, R.~F., et al.\ 2005, \aj, 129, 578. doi:10.1086/426915
  
\bibitem[Baskin \& Laor(2005)]{baskin2005} Baskin, A. \& Laor, A.\ 2005, \mnras, 356, 1029. doi:10.1111/j.1365-2966.2004.08525.x

\bibitem[Baumgartner et al.(2013)]{baumgartner2013} Baumgartner, W.~H., Tueller, J., Markwardt, C.~B., et al.\ 2013, \apjs, 207, 19. doi:10.1088/0067-0049/207/2/19

\bibitem[Becker et al.(1995)]{becker1995} Becker, R.~H., White, R.~L., \& Helfand, D.~J.\ 1995, \apj, 450, 559. doi:10.1086/176166

\bibitem[Bentz et al.(2009)]{bentz2009} Bentz, M.~C., Peterson, B.~M., Netzer, H., et al.\ 2009, \apj, 697, 160. doi:10.1088/0004-637X/697/1/160

\bibitem[Bentz et al.(2013)]{bentz2013} Bentz, M.~C., Denney, K.~D., Grier, C.~J., et al.\ 2013, \apj, 767, 149. doi:10.1088/0004-637X/767/2/149

\bibitem[Bentz \& Katz(2015)]{bentz2015} Bentz, M.~C. \& Katz, S.\ 2015, \pasp, 127, 67. doi:10.1086/679601

 \bibitem[Bertin \& Arnouts(1996)]{bertin1996} Bertin, E. \& Arnouts, S.\ 1996, \aaps, 117, 393. doi:10.1051/aas:1996164

\bibitem[Bird et al.(2010)]{bird2010} Bird, A.~J., Bazzano, A., Bassani, L., et al.\ 2010, \apjs, 186, 1. doi:10.1088/0067-0049/186/1/1

\bibitem[Blanton et al.(2017)]{blanton2017} Blanton, M.~R., Bershady, M.~A., Abolfathi, B., et al.\ 2017, \aj, 154, 28. doi:10.3847/1538-3881/aa7567

\bibitem[Bogdan et al.(2023)]{bogdan2023} Bogdan, A., Goulding, A., Natarajan, P., et al.\ 2023, arXiv:2305.15458. doi:10.48550/arXiv.2305.15458

\bibitem[Bonzini et al.(2013)]{bonzini2013} Bonzini, M., Padovani, P., Mainieri, V., et al.\ 2013, \mnras, 436, 3759. doi:10.1093/mnras/stt1879

\bibitem[Boroson \& Green(1992)]{boroson1992} Boroson, T.~A. \& Green, R.~F.\ 1992, \apjs, 80, 109. doi:10.1086/191661

\bibitem[Brandt \& Hasinger(2005)]{brandt2005} Brandt, W.~N. \& Hasinger, G.\ 2005, \araa, 43, 827. doi:10.1146/annurev.astro.43.051804.102213

\bibitem[Brunner et al.(2022)]{brunner2022} Brunner, H., Liu, T., Lamer, G., et al.\ 2022, \aap, 661, A1. doi:10.1051/0004-6361/202141266

\bibitem[Brusa et al.(2010)]{brusa2010} Brusa, M., Civano, F., Comastri, A., et al.\ 2010, \apj, 716, 348. doi:10.1088/0004-637X/716/1/348

\bibitem[Buchner et al.(2015)]{buchner2015} Buchner, J., Georgakakis, A., Nandra, K., et al.\ 2015, \apj, 802, 89. doi:10.1088/0004-637X/802/2/89

\bibitem[Bunker et al.(2023)]{bunker2023} Bunker, A.~J., Cameron, A.~J., Curtis-Lake, E., et al.\ 2023, arXiv:2306.02467. doi:10.48550/arXiv.2306.02467

\bibitem[Cappelluti et al.(2009)]{cappelluti2009} Cappelluti, N., Brusa, M., Hasinger, G., et al.\ 2009, \aap, 497, 635. doi:10.1051/0004-6361/200810794

\bibitem[Cann et al.(2020)]{cann2020} Cann, J.~M., Satyapal, S., Bohn, T., et al.\ 2020, \apj, 895, 147. doi:10.3847/1538-4357/ab8b64

\bibitem[Casali et al.(2007)]{casali2007} Casali, M., Adamson, A., Alves de Oliveira, C., et al.\ 2007, \aap, 467, 777. doi:10.1051/0004-6361:20066514

\bibitem[Chen et al.(2018)]{chen2018} Chen, C.-T.~J., Brandt, W.~N., Luo, B., et al.\ 2018, \mnras, 478, 2132. doi:10.1093/mnras/sty1036

\bibitem[Civano et al.(2016)]{civano2016} Civano, F., Marchesi, S., Comastri, A., et al.\ 2016, \apj, 819, 62. doi:10.3847/0004-637X/819/1/62

\bibitem[Coil et al.(2011)]{coil2011} Coil, A.~L., Blanton, M.~R., Burles, S.~M., et al.\ 2011, \apj, 741, 8. doi:10.1088/0004-637X/741/1/8

\bibitem[Coleman et al.(2022)]{coleman2022} Coleman, B., Kirkpatrick, A., Cooke, K.~C., et al.\ 2022, \mnras, 515, 82. doi:10.1093/mnras/stac1679

\bibitem[Comastri et al.(2002)]{comastri2002} Comastri, A., Mignoli, M., Ciliegi, P., et al.\ 2002, \apj, 571, 771. doi:10.1086/340016

\bibitem[Croom et al.(2009)]{croom2009} Croom, S.~M., Richards, G.~T., Shanks, T., et al.\ 2009, \mnras, 392, 19. doi:10.1111/j.1365-2966.2008.14052.x

\bibitem[Curtis-Lake et al.(2023)]{curtis-lake2023} Curtis-Lake, E., Carniani, S., Cameron, A., et al.\ 2023, Nature Astronomy, 7, 622. doi:10.1038/s41550-023-01918-w

\bibitem[Cushing et al.(2004)]{cushing2004} Cushing, M.~C., Vacca, W.~D., \& Rayner, J.~T.\ 2004, \pasp, 116, 362. doi:10.1086/382907

\bibitem[Cutri et al.(2021)]{allwise} Cutri, R.~M., Wright, E.~L., Conrow, T., et al.\ 2021, VizieR Online Data Catalog, II/328

\bibitem[Dawson et al.(2016)]{dawson2016} Dawson, K.~S., Kneib, J.-P., Percival, W.~J., et al.\ 2016, \aj, 151, 44. doi:10.3847/0004-6256/151/2/44
  
\bibitem[Denney et al.(2010)]{denney2010} Denney, K.~D., Peterson, B.~M., Pogge, R.~W., et al.\ 2010, \apj, 721, 715. doi:10.1088/0004-637X/721/1/715

\bibitem[Denney(2012)]{denney2012} Denney, K.~D.\ 2012, \apj, 759, 44. doi:10.1088/0004-637X/759/1/44

\bibitem[De Rosa et al.(2015)]{derosa2015} De Rosa, G., Peterson, B.~M., Ely, J., et al.\ 2015, \apj, 806, 128. doi:10.1088/0004-637X/806/1/128

\bibitem[Dewdney et al.(2009)]{dewdney2009} Dewdney, P.~E., Hall, P.~J., Schilizzi, R.~T., et al.\ 2009, IEEE Proceedings, 97, 1482. doi:10.1109/JPROC.2009.2021005
  
\bibitem[Dietrich et al.(2002)]{dietrich2002} Dietrich, M., Hamann, F., Shields, J.~C., et al.\ 2002, \apj, 581, 912. doi:10.1086/344410

\bibitem[Donley et al.(2012)]{donley2012} Donley, J.~L., Koekemoer, A.~M., Brusa, M., et al.\ 2012, \apj, 748, 142. doi:10.1088/0004-637X/748/2/142

\bibitem[Drinkwater et al.(2010)]{drinkwater2010} Drinkwater, M.~J., Jurek, R.~J., Blake, C., et al.\ 2010, \mnras, 401, 1429. doi:10.1111/j.1365-2966.2009.15754.x

\bibitem[Duras et al.(2020)]{duras2020} Duras, F., Bongiorno, A., Ricci, F., et al.\ 2020, \aap, 636, A73. doi:10.1051/0004-6361/201936817

\bibitem[Eisenstein et al.(2023)]{eisenstein2023} Eisenstein, D.~J., Willott, C., Alberts, S., et al.\ 2023, arXiv:2306.02465. doi:10.48550/arXiv.2306.02465

\bibitem[Elias et al.(1998)]{elias1998} Elias, J.~H., Vukobratovich, D., Andrew, J.~R., et al.\ 1998, \procspie, 3354, 555. doi:10.1117/12.317281

\bibitem[Evans et al.(2010)]{evans2010} Evans, I.~N., Primini, F.~A., Glotfelty, K.~J., et al.\ 2010, \apjs, 189, 37. doi:10.1088/0067-0049/189/1/37

\bibitem[Faber et al.(2003)]{faber2003} Faber, S.~M., Phillips, A.~C., Kibrick, R.~I., et al.\ 2003, \procspie, 4841, 1657. doi:10.1117/12.460346

\bibitem[Finkelstein et al.(2023)]{finkelstein2023} Finkelstein, S.~L., Bagley, M.~B., Ferguson, H.~C., et al.\ 2023, \apjl, 946, L13. doi:10.3847/2041-8213/acade4

\bibitem[Fliri \& Trujillo(2016)]{fliri2016} Fliri, J. \& Trujillo, I.\ 2016, \mnras, 456, 1359. doi:10.1093/mnras/stv2686

\bibitem[Frieman et al.(2008)]{frieman2008} Frieman, J.~A., Bassett, B., Becker, A., et al.\ 2008, \aj, 135, 338. doi:10.1088/0004-6256/135/1/338
  
\bibitem[Fujimoto et al.(2023)]{fujimoto2023} Fujimoto, S., Arrabal Haro, P., Dickinson, M., et al.\ 2023, \apjl, 949, L25. doi:10.3847/2041-8213/acd2d9

\bibitem[Garilli et al.(2008)]{garilli2008} Garilli, B., Le F{\`e}vre, O., Guzzo, L., et al.\ 2008, \aap, 486, 683. doi:10.1051/0004-6361:20078878
  
\bibitem[Ghosh et al.(2022)]{ghosh2022} Ghosh, A., Urry, C.~M., Rau, A., et al.\ 2022, \apj, 935, 138. doi:10.3847/1538-4357/ac7f9e

\bibitem[Giacconi et al.(2002)]{giacconi2002} Giacconi, R., Zirm, A., Wang, J., et al.\ 2002, \apjs, 139, 369. doi:10.1086/338927

\bibitem[Giavalisco et al.(2004)]{giavalisco2004} Giavalisco, M., Ferguson, H.~C., Koekemoer, A.~M., et al.\ 2004, \apjl, 600, L93. doi:10.1086/379232

\bibitem[Glikman et al.(2004)]{glikman2004} Glikman, E., Gregg, M.~D., Lacy, M., et al.\ 2004, \apj, 607, 60. doi:10.1086/383305

\bibitem[Glikman et al.(2007)]{glikman2007} Glikman, E., Helfand, D.~J., White, R.~L., et al.\ 2007, \apj, 667, 673. doi:10.1086/521073

\bibitem[Glikman et al.(2013)]{glikman2013} Glikman, E., Urrutia, T., Lacy, M., et al.\ 2013, \apj, 778, 127. doi:10.1088/0004-637X/778/2/127

\bibitem[Glikman et al.(2018)]{glikman2018} Glikman, E., Lacy, M., LaMassa, S., et al.\ 2018, \apj, 861, 37. doi:10.3847/1538-4357/aac5d8

\bibitem[Glikman et al.(2022)]{glikman2022} Glikman, E., Lacy, M., LaMassa, S., et al.\ 2022, \apj, 934, 119. doi:10.3847/1538-4357/ac6bee
  

\bibitem[Goulding et al.(2023)]{goulding2023} Goulding, A.~D., Greene, J.~E., Setton, D.~J., et al.\ 2023, \apjl, 955, L24. doi:10.3847/2041-8213/acf7c5

\bibitem[Greene \& Ho(2005)]{greene2005} Greene, J.~E. \& Ho, L.~C.\ 2005, \apj, 630, 122. doi:10.1086/431897

\bibitem[Greene et al.(2010)]{greene2010} Greene, J.~E., Peng, C.~Y., \& Ludwig, R.~R.\ 2010, \apj, 709, 937. doi:10.1088/0004-637X/709/2/937

\bibitem[Grier et al.(2012)]{grier2012} Grier, C.~J., Peterson, B.~M., Pogge, R.~W., et al.\ 2012, \apj, 755, 60. doi:10.1088/0004-637X/755/1/60

\bibitem[Grier et al.(2017)]{grier2017} Grier, C.~J., Trump, J.~R., Shen, Y., et al.\ 2017, \apj, 851, 21. doi:10.3847/1538-4357/aa98dc

 \bibitem[Grier et al.(2019)]{grier2019} Grier, C.~J., Shen, Y., Horne, K., et al.\ 2019, \apj, 887, 38. doi:10.3847/1538-4357/ab4ea5

\bibitem[Gunn et al.(2006)]{gunn2006} Gunn, J.~E., Siegmund, W.~A., Mannery, E.~J., et al.\ 2006, \aj, 131, 2332. doi:10.1086/500975

\bibitem[Guo et al.(2018)]{guo2018} Guo, H., Shen, Y., \& Wang, S.\ 2018, Astrophysics Source Code Library. ascl:1809.008

\bibitem[Hao et al.(2005)]{hao2005} Hao, L., Strauss, M.~A., Tremonti, C.~A., et al.\ 2005, \aj, 129, 1783. doi:10.1086/428485
  
\bibitem[Hasinger et al.(2007)]{hasinger2007} Hasinger, G., Cappelluti, N., Brunner, H., et al.\ 2007, \apjs, 172, 29. doi:10.1086/516576

\bibitem[Helfand et al.(2015)]{helfand2015} Helfand, D.~J., White, R.~L., \& Becker, R.~H.\ 2015, \apj, 801, 26. doi:10.1088/0004-637X/801/1/26

\bibitem[Herter et al.(2008)]{herter2008} Herter, T.~L., Henderson, C.~P., Wilson, J.~C., et al.\ 2008, \procspie, 7014, 70140X. doi:10.1117/12.789660

\bibitem[Hewett et al.(2006)]{hewett2006} Hewett, P.~C., Warren, S.~J., Leggett, S.~K., et al.\ 2006, \mnras, 367, 454. doi:10.1111/j.1365-2966.2005.09969.x

\bibitem[Hickox \& Alexander(2018)]{hickox2018} Hickox, R.~C. \& Alexander, D.~M.\ 2018, \araa, 56, 625. doi:10.1146/annurev-astro-081817-051803

\bibitem[Hodge et al.(2011)]{hodge2011} Hodge, J.~A., Becker, R.~H., White, R.~L., et al.\ 2011, \aj, 142, 3. doi:10.1088/0004-6256/142/1/3

\bibitem[H{\"o}nig \& Beckert(2007)]{hoenig2007} H{\"o}nig, S.~F. \& Beckert, T.\ 2007, \mnras, 380, 1172. doi:10.1111/j.1365-2966.2007.12157.x
  
\bibitem[Hopkins et al.(2006)]{hopkins2006} Hopkins, P.~F., Hernquist, L., Cox, T.~J., et al.\ 2006, \apjs, 163, 1. doi:10.1086/499298

\bibitem[Hopkins \& Hernquist(2009)]{hopkins2009} Hopkins, P.~F. \& Hernquist, L.\ 2009, \apj, 698, 1550. doi:10.1088/0004-637X/698/2/1550

\bibitem[Hornschemeier et al.(2005)]{hornschemeier2005} Hornschemeier, A.~E., Heckman, T.~M., Ptak, A.~F., et al.\ 2005, \aj, 129, 86. doi:10.1086/426371

\bibitem[Hviding et al.(2022b)]{hviding2022} Hviding, R.~E., Hainline, K.~N., Rieke, M., et al.\ 2022, \aj, 163, 224. doi:10.3847/1538-3881/ac5e33

\bibitem[Hviding(2022a)]{gelato} Hviding, R.~E.\ 2022, Zenodo

\bibitem[Ichikawa et al.(2017)]{ichikawa2017} Ichikawa, K., Ricci, C., Ueda, Y., et al.\ 2017, \apj, 835, 74. doi:10.3847/1538-4357/835/1/74

\bibitem[Jiang et al.(2006)]{jiang2006} Jiang, L., Fan, X., Cool, R.~J., et al.\ 2006, \aj, 131, 2788. doi:10.1086/503745

\bibitem[Jiang et al.(2014)]{jiang2014} Jiang, L., Fan, X., Bian, F., et al.\ 2014, \apjs, 213, 12. doi:10.1088/0067-0049/213/1/12

\bibitem[Jiang et al.(2022)]{jiang2022} Jiang, L., Ning, Y., Fan, X., et al.\ 2022, Nature Astronomy, 6, 850. doi:10.1038/s41550-022-01708-w


\bibitem[Jonas \& MeerKAT Team(2016)]{jonas2016} Jonas, J. \& MeerKAT Team\ 2016, MeerKAT Science: On the Pathway to the SKA, 1. doi:10.22323/1.277.0001

\bibitem[Jones et al.(2004)]{jones2004} Jones, D.~H., Saunders, W., Colless, M., et al.\ 2004, \mnras, 355, 747. doi:10.1111/j.1365-2966.2004.08353.x

\bibitem[Jones et al.(2009)]{jones2009} Jones, D.~H., Read, M.~A., Saunders, W., et al.\ 2009, \mnras, 399, 683. doi:10.1111/j.1365-2966.2009.15338.x


\bibitem[Kaspi et al.(2007)]{kaspi2007} Kaspi, S., Brandt, W.~N., Maoz, D., et al.\ 2007, \apj, 659, 997. doi:10.1086/512094

\bibitem[Kassis et al.(2018)]{kassis2018} Kassis, M., Chan, D., Kwok, S., et al.\ 2018, \procspie, 10702, 1070207. doi:10.1117/12.2312316

\bibitem[Kirkpatrick et al.(2020)]{kirkpatrick2020} Kirkpatrick, A., Urry, C.~M., Brewster, J., et al.\ 2020, \apj, 900, 5. doi:10.3847/1538-4357/aba358
  
\bibitem[Kollmeier et al.(2006)]{kollmeier2006} Kollmeier, J.~A., Onken, C.~A., Kochanek, C.~S., et al.\ 2006, \apj, 648, 128. doi:10.1086/505646

\bibitem[Koss et al.(2016)]{koss2016} Koss, M.~J., Assef, R., Balokovi{\'c}, M., et al.\ 2016, \apj, 825, 85. doi:10.3847/0004-637X/825/2/85

\bibitem[Koss et al.(2017)]{koss2017} Koss, M., Trakhtenbrot, B., Ricci, C., et al.\ 2017, \apj, 850, 74. doi:10.3847/1538-4357/aa8ec9

\bibitem[LaMassa et al.(2013a)]{lamassa2013a} LaMassa, S.~M., Urry, C.~M., Glikman, E., et al.\ 2013, \mnras, 432, 1351. doi:10.1093/mnras/stt553

\bibitem[LaMassa et al.(2013b)]{lamassa2013b} LaMassa, S.~M., Urry, C.~M., Cappelluti, N., et al.\ 2013, \mnras, 436, 3581. doi:10.1093/mnras/stt1837

\bibitem[LaMassa et al.(2016a)]{lamassa2016a} LaMassa, S.~M., Urry, C.~M., Cappelluti, N., et al.\ 2016, \apj, 817, 172. doi:10.3847/0004-637X/817/2/172

\bibitem[LaMassa et al.(2016b)]{lamassa2016b} LaMassa, S.~M., Civano, F., Brusa, M., et al.\ 2016, \apj, 818, 88. doi:10.3847/0004-637X/818/1/88

\bibitem[LaMassa et al.(2017)]{lamassa2017} LaMassa, S.~M., Glikman, E., Brusa, M., et al.\ 2017, \apj, 847, 100. doi:10.3847/1538-4357/aa87b5

\bibitem[LaMassa et al.(2019)]{lamassa2019} LaMassa, S.~M., Georgakakis, A., Vivek, M., et al.\ 2019, \apj, 876, 50. doi:10.3847/1538-4357/ab108b

\bibitem[LaMassa et al.(2023a)]{lamassa2023a} LaMassa, S.~M., Yaqoob, T., Tzanavaris, P., et al.\ 2023a, \apj, 944, 152. doi:10.3847/1538-4357/acb3bb

\bibitem[LaMassa et al.(2024)]{lamassa2024} LaMassa, S.~M., Farrow, I., Urry, C. M. et al.\ 2024, {\it submitted to AAS Journals}

\bibitem[Lawrence(1991)]{lawrence1991} Lawrence, A.\ 1991, \mnras, 252, 586. doi:10.1093/mnras/252.4.586

\bibitem[Lawrence et al.(2007)]{lawrence2007} Lawrence, A., Warren, S.~J., Almaini, O., et al.\ 2007, \mnras, 379, 1599. doi:10.1111/j.1365-2966.2007.12040.x

\bibitem[Le F{\`e}vre et al.(2004)]{lefevre2004} Le F{\`e}vre, O., Vettolani, G., Paltani, S., et al.\ 2004, \aap, 428, 1043. doi:10.1051/0004-6361:20048072

\bibitem[Le F{\`e}vre et al.(2005)]{lefevre2005} Le F{\`e}vre, O., Vettolani, G., Garilli, B., et al.\ 2005, \aap, 439, 845. doi:10.1051/0004-6361:20041960

\bibitem[Le F{\`e}vre et al.(2013)]{lefevre2013} Le F{\`e}vre, O., Cassata, P., Cucciati, O., et al.\ 2013, \aap, 559, A14. doi:10.1051/0004-6361/201322179

\bibitem[Lehmer et al.(2005)]{lehmer2005} Lehmer, B.~D., Brandt, W.~N., Alexander, D.~M., et al.\ 2005, \apjs, 161, 21. doi:10.1086/444590

\bibitem[Lira et al.(2018)]{lira2018} Lira, P., Kaspi, S., Netzer, H., et al.\ 2018, \apj, 865, 56. doi:10.3847/1538-4357/aada45

\bibitem[Liu et al.(2016)]{liu2016} Liu, Z., Merloni, A., Georgakakis, A., et al.\ 2016, \mnras, 459, 1602. doi:10.1093/mnras/stw753

\bibitem[Liu et al.(2017)]{liu2017} Liu, T., Tozzi, P., Wang, J.-X., et al.\ 2017, \apjs, 232, 8. doi:10.3847/1538-4365/aa7847

\bibitem[Liu et al.(2022)]{liu2022} Liu, T., Merloni, A., Comparat, J., et al.\ 2022, \aap, 661, A27. doi:10.1051/0004-6361/202141178


\bibitem[Luo et al.(2008)]{luo2008} Luo, B., Bauer, F.~E., Brandt, W.~N., et al.\ 2008, \apjs, 179, 19. doi:10.1086/591248  

\bibitem[Luo et al.(2017)]{luo2017} Luo, B., Brandt, W.~N., Xue, Y.~Q., et al.\ 2017, \apjs, 228, 2. doi:10.3847/1538-4365/228/1/2
  
\bibitem[Lusso et al.(2012)]{lusso2012} Lusso, E., Comastri, A., Simmons, B.~D., et al.\ 2012, \mnras, 425, 623. doi:10.1111/j.1365-2966.2012.21513.x

\bibitem[Lusso et al.(2013)]{lusso2013} Lusso, E., Hennawi, J.~F., Comastri, A., et al.\ 2013, \apj, 777, 86. doi:10.1088/0004-637X/777/2/86

\bibitem[Lyke et al.(2020)]{lyke2020} Lyke, B.~W., Higley, A.~N., McLane, J.~N., et al.\ 2020, \apjs, 250, 8. doi:10.3847/1538-4365/aba623

\bibitem[McKernan \& Yaqoob(1998)]{mckernan1998} McKernan, B. \& Yaqoob, T.\ 1998, \apjl, 501, L29. doi:10.1086/311457

\bibitem[McMahon et al.(2013)]{mcmahon2013} McMahon, R.~G., Banerji, M., Gonzalez, E., et al.\ 2013, The Messenger, 154, 35

\bibitem[Madau \& Haardt(2015)]{madau2015} Madau, P. \& Haardt, F.\ 2015, \apjl, 813, L8. doi:10.1088/2041-8205/813/1/L8

\bibitem[Maiolino et al.(2007)]{maiolino2007} Maiolino, R., Shemmer, O., Imanishi, M., et al.\ 2007, \aap, 468, 979. doi:10.1051/0004-6361:20077252

\bibitem[Malizia et al.(2012)]{malizia2012} Malizia, A., Bassani, L., Bazzano, A., et al.\ 2012, \mnras, 426, 1750. doi:10.1111/j.1365-2966.2012.21755.x

\bibitem[Marchesi et al.(2016a)]{marchesi2016a} Marchesi, S., Civano, F., Elvis, M., et al.\ 2016, \apj, 817, 34. doi:10.3847/0004-637X/817/1/34

\bibitem[Marchesi et al.(2016b)]{marchesi2016b} Marchesi, S., Civano, F., Salvato, M., et al.\ 2016, \apj, 827, 150. doi:10.3847/0004-637X/827/2/150

\bibitem[Marinucci et al.(2016)]{marinucci2016} Marinucci, A., Bianchi, S., Matt, G., et al.\ 2016, \mnras, 456, L94. doi:10.1093/mnrasl/slv178

\bibitem[Marziani et al.(1996)]{marziani1996} Marziani, P., Sulentic, J.~W., Dultzin-Hacyan, D., et al.\ 1996, \apjs, 104, 37. doi:10.1086/192291

\bibitem[Masini et al.(2020)]{masini2020} Masini, A., Hickox, R.~C., Carroll, C.~M., et al.\ 2020, \apjs, 251, 2. doi:10.3847/1538-4365/abb607

\bibitem[Mateos et al.(2012)]{mateos2012} Mateos, S., Alonso-Herrero, A., Carrera, F.~J., et al.\ 2012, \mnras, 426, 3271. doi:10.1111/j.1365-2966.2012.21843.x

\bibitem[Mateos et al.(2017)]{mateos2017} Mateos, S., Carrera, F.~J., Barcons, X., et al.\ 2017, \apjl, 841, L18. doi:10.3847/2041-8213/aa7268

\bibitem[McLean et al.(1998)]{mclean1998} McLean, I.~S., Becklin, E.~E., Bendiksen, O., et al.\ 1998, \procspie, 3354, 566. doi:10.1117/12.317283

\bibitem[McLure \& Jarvis(2002)]{mclure2002} McLure, R.~J. \& Jarvis, M.~J.\ 2002, \mnras, 337, 109. doi:10.1046/j.1365-8711.2002.05871.x

\bibitem[Medvedev et al.(2021)]{medvedev2021} Medvedev, P., Gilfanov, M., Sazonov, S., et al.\ 2021, \mnras, 504, 576. doi:10.1093/mnras/stab773
  
\bibitem[Medvedev et al.(2020)]{medvedev2020} Medvedev, P., Sazonov, S., Gilfanov, M., et al.\ 2020, \mnras, 497, 1842. doi:10.1093/mnras/staa2051
  
\bibitem[Mendez et al.(2013)]{mendez2013} Mendez, A.~J., Coil, A.~L., Aird, J., et al.\ 2013, \apj, 770, 40. doi:10.1088/0004-637X/770/1/40

\bibitem[Menzel et al.(2016)]{menzel2016} Menzel, M.-L., Merloni, A., Georgakakis, A., et al.\ 2016, \mnras, 457, 110. doi:10.1093/mnras/stv2749

\bibitem[Merloni et al.(2024)]{merloni2024} Merloni, A., Lamer, G., Liu, T., et al.\ 2024, \aap, 682, A34. doi:10.1051/0004-6361/202347165
  
\bibitem[Merloni et al.(2014)]{merloni2014} Merloni, A., Bongiorno, A., Brusa, M., et al.\ 2014, \mnras, 437, 3550. doi:10.1093/mnras/stt2149

\bibitem[Morrissey et al.(2007)]{morrissey2007} Morrissey, P., Conrow, T., Barlow, T.~A., et al.\ 2007, \apjs, 173, 682. doi:10.1086/520512

\bibitem[Netzer et al.(2007)]{netzer2007} Netzer, H., Lira, P., Trakhtenbrot, B., et al.\ 2007, \apj, 671, 1256. doi:10.1086/523035

\bibitem[Netzer(2009)]{netzer2009} Netzer, H.\ 2009, \mnras, 399, 1907. doi:10.1111/j.1365-2966.2009.15434.x

\bibitem[Netzer(2015)]{netzer2015} Netzer, H.\ 2015, \araa, 53, 365. doi:10.1146/annurev-astro-082214-122302

\bibitem[Newman et al.(2013)]{newman2013} Newman, J.~A., Cooper, M.~C., Davis, M., et al.\ 2013, \apjs, 208, 5. doi:10.1088/0067-0049/208/1/5

\bibitem[Ni et al.(2021)]{ni2021} Ni, Q., Brandt, W.~N., Chen, C.-T., et al.\ 2021, \apjs, 256, 21. doi:10.3847/1538-4365/ac0dc6

\bibitem[Oh et al.(2018)]{oh2018} Oh, K., Koss, M., Markwardt, C.~B., et al.\ 2018, \apjs, 235, 4. doi:10.3847/1538-4365/aaa7fd

\bibitem[Oke \& Gunn(1982)]{oke1982} Oke, J.~B. \& Gunn, J.~E.\ 1982, \pasp, 94, 586. doi:10.1086/131027

\bibitem[Oke et al.(1995)]{oke1995} Oke, J.~B., Cohen, J.~G., Carr, M., et al.\ 1995, \pasp, 107, 375. doi:10.1086/133562

\bibitem[Osterbrock \& Ferland(2006)]{osterbrock2006} Osterbrock, D.~E. \& Ferland, G.~J.\ 2006, Astrophysics of gaseous nebulae and active galactic nuclei, 2nd. ed. by D.E. Osterbrock and G.J. Ferland. Sausalito, CA: University Science Books, 2006

\bibitem[Papovich et al.(2016)]{papovich2016} Papovich, C., Shipley, H.~V., Mehrtens, N., et al.\ 2016, \apjs, 224, 28. doi:10.3847/0067-0049/224/2/28

\bibitem[P{\^a}ris et al.(2017)]{dr12q} P{\^a}ris, I., Petitjean, P., Ross, N.~P., et al.\ 2017, \aap, 597, A79. doi:10.1051/0004-6361/201527999

\bibitem[P{\^a}ris et al.(2018)]{dr14q} P{\^a}ris, I., Petitjean, P., Aubourg, {\'E}., et al.\ 2018, \aap, 613, A51. doi:10.1051/0004-6361/201732445

\bibitem[Peca et al.(2023)]{peca2023} Peca, A., Cappelluti, N., Urry, C.~M., et al.\ 2023, \apj, 943, 162. doi:10.3847/1538-4357/acac28

\bibitem[Peterson et al.(2002)]{peterson2002} Peterson, B.~M., Berlind, P., Bertram, R., et al.\ 2002, \apj, 581, 197. doi:10.1086/344197

\bibitem[Peterson et al.(2004)]{peterson2004} Peterson, B.~M., Ferrarese, L., Gilbert, K.~M., et al.\ 2004, \apj, 613, 682. doi:10.1086/423269

\bibitem[Peterson(2014)]{peterson2014} Peterson, B.~M.\ 2014, \ssr, 183, 253. doi:10.1007/s11214-013-9987-4

\bibitem[Pierre et al.(2016)]{pierre2016} Pierre, M., Pacaud, F., Adami, C., et al.\ 2016, \aap, 592, A1. doi:10.1051/0004-6361/201526766

\bibitem[Predehl et al.(2021)]{predehl2021} Predehl, P., Andritschke, R., Arefiev, V., et al.\ 2021, \aap, 647, A1. doi:10.1051/0004-6361/202039313

\bibitem[Prochaska et al.(2020a)]{prochaska2020a} Prochaska, J., Hennawi, J., Westfall, K., et al.\ 2020, The Journal of Open Source Software, 5, 2308. doi:10.21105/joss.02308

\bibitem[Prochaska et al.(2020b)]{prochaska2020b} Prochaska, J.~X., Hennawi, J., Cooke, R., et al.\ 2020, Zenodo

\bibitem[Rahmer et al.(2012)]{rahmer2012} Rahmer, G., Smith, R.~M., Bui, K., et al.\ 2012, \procspie, 8446, 84462Z. doi:10.1117/12.926687

\bibitem[Richards et al.(2002)]{richards2002} Richards, G.~T., Fan, X., Newberg, H.~J., et al.\ 2002, \aj, 123, 2945. doi:10.1086/340187

\bibitem[Rieke et al.(2023)]{rieke2023} Rieke, M.~J., Robertson, B.~E., Tacchella, S., et al.\ 2023, arXiv:2306.02466. doi:10.48550/arXiv.2306.02466

\bibitem[Risaliti et al.(2009)]{risaliti2009} Risaliti, G., Salvati, M., Elvis, M., et al.\ 2009, \mnras, 393, L1. doi:10.1111/j.1745-3933.2008.00580.x
  
\bibitem[Rockosi et al.(2010)]{rockosi2010} Rockosi, C., Stover, R., Kibrick, R., et al.\ 2010, \procspie, 7735, 77350R. doi:10.1117/12.856818

\bibitem[Ross et al.(2012)]{ross2012} Ross, N.~P., Myers, A.~D., Sheldon, E.~S., et al.\ 2012, \apjs, 199, 3. doi:10.1088/0067-0049/199/1/3

\bibitem[Runnoe et al.(2012)]{runnoe2012} Runnoe, J.~C., Brotherton, M.~S., \& Shang, Z.\ 2012, \mnras, 422, 478. doi:10.1111/j.1365-2966.2012.20620.x

\bibitem[Salvato et al.(2022)]{salvato2022} Salvato, M., Wolf, J., Dwelly, T., et al.\ 2022, \aap, 661, A3. doi:10.1051/0004-6361/202141631

\bibitem[Sanders et al.(1988)]{sanders1988} Sanders, D.~B., Soifer, B.~T., Elias, J.~H., et al.\ 1988, \apjl, 328, L35. doi:10.1086/185155

\bibitem[Schneider et al.(2010)]{dr7q} Schneider, D.~P., Richards, G.~T., Hall, P.~B., et al.\ 2010, \aj, 139, 2360. doi:10.1088/0004-6256/139/6/2360

 \bibitem[Schulze et al.(2018)]{schulze2018} Schulze, A., Silverman, J.~D., Kashino, D., et al.\ 2018, \apjs, 239, 22. doi:10.3847/1538-4365/aae82f

\bibitem[Shen et al.(2008)]{shen2008} Shen, Y., Greene, J.~E., Strauss, M.~A., et al.\ 2008, \apj, 680, 169. doi:10.1086/587475

\bibitem[Shen et al.(2011)]{shen2011} Shen, Y., Richards, G.~T., Strauss, M.~A., et al.\ 2011, \apjs, 194, 45. doi:10.1088/0067-0049/194/2/45

\bibitem[Shen \& Liu(2012)]{shen2012} Shen, Y. \& Liu, X.\ 2012, \apj, 753, 125. doi:10.1088/0004-637X/753/2/125

\bibitem[Shen et al.(2024)]{shen2024} Shen, Y., Grier, C.~J., Horne, K., et al.\ 2024, \apjs, 272, 26. doi:10.3847/1538-4365/ad3936

\bibitem[Smee et al.(2013)]{smee2013} Smee, S.~A., Gunn, J.~E., Uomoto, A., et al.\ 2013, \aj, 146, 32. doi:10.1088/0004-6256/146/2/32

\bibitem[Soltan(1982)]{soltan1982} Soltan, A.\ 1982, \mnras, 200, 115. doi:10.1093/mnras/200.1.115

\bibitem[Sutherland \& Saunders(1992)]{sutherland1992} Sutherland, W. \& Saunders, W.\ 1992, \mnras, 259, 413. doi:10.1093/mnras/259.3.413

 \bibitem[Temple et al.(2023)]{temple2023} Temple, M.~J., Matthews, J.~H., Hewett, P.~C., et al.\ 2023, \mnras, 523, 646. doi:10.1093/mnras/stad1448


\bibitem[Tian et al.(2023)]{tian2023} Tian, C., Urry, C.~M., Ghosh, A., et al.\ 2023, \apj, 944, 124. doi:10.3847/1538-4357/acad79

\bibitem[Timlin et al.(2016)]{timlin2016} Timlin, J.~D., Ross, N.~P., Richards, G.~T., et al.\ 2016, \apjs, 225, 1. doi:10.3847/0067-0049/225/1/1

\bibitem[Trakhtenbrot \& Netzer(2012)]{trakhtenbrot2012} Trakhtenbrot, B. \& Netzer, H.\ 2012, \mnras, 427, 3081. doi:10.1111/j.1365-2966.2012.22056.x

\bibitem[Treister et al.(2012)]{treister2012} Treister, E., Schawinski, K., Urry, C.~M., et al.\ 2012, \apjl, 758, L39. doi:10.1088/2041-8205/758/2/L39

\bibitem[Ueda et al.(2014)]{ueda2014} Ueda, Y., Akiyama, M., Hasinger, G., et al.\ 2014, \apj, 786, 104. doi:10.1088/0004-637X/786/2/104

 \bibitem[Urrutia et al.(2009)]{urrutia2009} Urrutia, T., Becker, R.~H., White, R.~L., et al.\ 2009, \apj, 698, 1095. doi:10.1088/0004-637X/698/2/1095

\bibitem[Urry \& Padovani(1995)]{urry1995} Urry, C.~M. \& Padovani, P.\ 1995, \pasp, 107, 803. doi:10.1086/133630

\bibitem[Vazdekis et al.(2016)]{vazdekis2016} Vazdekis, A., Koleva, M., Ricciardelli, E., et al.\ 2016, \mnras, 463, 3409. doi:10.1093/mnras/stw2231
  
\bibitem[Vestergaard \& Wilkes(2001)]{vestergaard2001} Vestergaard, M. \& Wilkes, B.~J.\ 2001, \apjs, 134, 1. doi:10.1086/320357

  \bibitem[Vestergaard \& Peterson(2006)]{vestergaard2006} Vestergaard, M. \& Peterson, B.~M.\ 2006, \apj, 641, 689. doi:10.1086/500572

\bibitem[Viero et al.(2014)]{viero2014} Viero, M.~P., Asboth, V., Roseboom, I.~G., et al.\ 2014, \apjs, 210, 22. doi:10.1088/0067-0049/210/2/22

\bibitem[Webster et al.(1995)]{webster} Webster, R.~L., Francis, P.~J., Petersont, B.~A., et al.\ 1995, \nat, 375, 469. doi:10.1038/375469a0

\bibitem[White et al.(1997)]{white1997} White, R.~L., Becker, R.~H., Helfand, D.~J., et al.\ 1997, \apj, 475, 479. doi:10.1086/303564

 \bibitem[White et al.(2000)]{white2000} White, R.~L., Becker, R.~H., Gregg, M.~D., et al.\ 2000, \apjs, 126, 133. doi:10.1086/313300


\bibitem[Wilson et al.(2004)]{wilson2004} Wilson, J.~C., Henderson, C.~P., Herter, T.~L., et al.\ 2004, \procspie, 5492, 1295. doi:10.1117/12.550925

\bibitem[Wolf et al.(2023)]{wolf2023} Wolf, J., Nandra, K., Salvato, M., et al.\ 2023, \aap, 669, A127. doi:10.1051/0004-6361/202244688

 \bibitem[Wolf et al.(2021)]{wolf2021} Wolf, J., Nandra, K., Salvato, M., et al.\ 2021, \aap, 647, A5. doi:10.1051/0004-6361/202039724

\bibitem[Wright et al.(2010)]{wright2010} Wright, E.~L., Eisenhardt, P.~R.~M., Mainzer, A.~K., et al.\ 2010, \aj, 140, 1868. doi:10.1088/0004-6256/140/6/1868

\bibitem[Xue et al.(2011)]{xue2011} Xue, Y.~Q., Luo, B., Brandt, W.~N., et al.\ 2011, \apjs, 195, 10. doi:10.1088/0067-0049/195/1/10

\bibitem[Yip et al.(2004)]{yip2004} Yip, C.~W., Connolly, A.~J., Szalay, A.~S., et al.\ 2004, \aj, 128, 585. doi:10.1086/422429

\bibitem[Yung et al.(2021)]{yung2021} Yung, L.~Y.~A., Somerville, R.~S., Finkelstein, S.~L., et al.\ 2021, \mnras, 508, 2706. doi:10.1093/mnras/stab2761
  
\end{thebibliography}
\end{document}